\documentclass[aps,prb]{revtex4-2}  

\usepackage[version=3]{mhchem}	
\usepackage{graphicx}  
\usepackage{nicefrac}  
\usepackage{siunitx} 	
\usepackage{amsmath,amssymb,amsfonts}   
\usepackage{hyperref}
\usepackage{import}
\usepackage{mathdesign}
\usepackage{bbold}
\usepackage{wasysym}
\usepackage{xtab}
\usepackage{xtab,afterpage}
\usepackage{lipsum}
\usepackage{multirow}
\usepackage{bm}
\usepackage[normalem]{ulem}
\usepackage{array}
\usepackage{graphicx}
\usepackage{cancel}
\usepackage{booktabs}
\usepackage{comment}

\usepackage{isomath}
\usepackage{bbm}
\usepackage{braket}

\usepackage{xr}
\newcommand{\hmn}[1]{
  \ensuremath{\begingroup\setupHMN #1\endgroup}%
}

\newcommand{\setupHMN}{%
  \doHMN{-}{\HMNoverline}%
  \doHMN{*}{\HMNminverse}%
    \doHMN{£}{\HMNninverse}%
  \doHMN{`}{\HMNminversep}%
  \doHMN{~}{\HMNminversea}%
  \doHMN{!}{\HMNminversemtilde}%
  \doHMN{i}{\infty}
}

\newcommand{\doHMN}[2]{%
  \begingroup\lccode`~=`#1
  \lowercase{\endgroup\let~}#2%
  \mathcode`#1="8000
}

\newcommand{\HMNminverse}[1]{\frac{#1}{m}}
\newcommand{\HMNninverse}[1]{\frac{#1}{n}}
\newcommand{\HMNminversep}[1]{\frac{#1}{m'}}
\newcommand{\HMNminversea}[1]{\frac{#1}{a}}
\newcommand{\HMNminversemtilde}[1]{\frac{#1}{\tilde{m}}}
\newcommand{\HMNoverline}[1]{\mkern1mu\overline{\mkern-1mu#1\mkern-1mu}\mkern1mu}

\newcommand{\mat}{\matrixsym}
\newcommand{\ten}{\tensorsym}

\newcommand{\irrep}{\textit{irrep}}
\newcommand{\irreps}{\textit{irreps}}

\newcommand{\tcell}[1]{\parbox[t]{4.0cm}{#1}}

\hypersetup{colorlinks=true,urlcolor=blue,citecolor=blue,linkcolor=blue,breaklinks}
\renewcommand{\vec}[1]{\mathbf{#1}}

\renewcommand{\arraystretch}{0.75}

\begin{document}

\title{Altermagnetism and generalised tensorial properties in the exchange multiplet  framework}

\author{Paolo G. Radaelli}
\affiliation{Clarendon Laboratory, Department of Physics, University of Oxford, Oxford, OX1 3PU, United Kingdom}
\email[Corresponding author: ]{p.g.radaelli@physics.ox.ac.uk}

\date{\today}

\begin{abstract}

A representation-theoretic framework for the calculation of time-reversal-odd tensorial properties of magnetic materials based on the Bertaut--Izyumov exchange-multiplet formalism is developed. In contrast to approaches based on distinct magnetic point groups or spin groups for different orientations of the magnetic order parameter, the exchange-multiplet framework permits the order parameter to be treated as a continuous variable. It is shown that, for constant-moment collinear structures, covariance imposes a simple set of symmetry constraints on the tensor intertwiners appearing in a Landau--Lifshitz expansion. This leads to canonical intertwiner forms that factorise the magnetic and crystallographic parts of the tensor construction. Tensor fields can be built to retain a Cartesian representation in the physical variables while acquiring a systematic dependence on the order-parameter direction through symmetry-adapted polynomial bases. Magnetic-point-group selection rules are recovered automatically for arbitrary order-parameter directions, while the framework also provides a natural separation of co-rotating (SOC-free) and non-co-rotating contributions. The approach is applicable to altermagnets and conventional antiferromagnets alike and is particularly suited to experiments and first-principles calculations in which the magnetic order parameter can be continuously varied.

 \end{abstract}

\maketitle

\section{Introduction}
In the past few years, the emergence of altermagnetism\cite{Smejkal2022, Smejkal2022b} has rekindled interest in the influence of magnetic anisotropy and the direction of the magnetic order parameter on the tensorial properties of materials.  In the absence of magnetic anisotropy, the spin dependence of tensorial properties follows the magnetic order parameter in a straightforward manner, whereas the dependence on other physical variables may exhibit highly non-trivial angular structure, as exemplified by the now familiar $d$-, $g$- and $i$-wave momentum textures.  When magnetic anisotropy is introduced, typically through spin--orbit coupling (SOC), tensorial properties acquire an additional angular dependence on the direction of the order parameter, which is generally described in terms of the corresponding magnetic point groups (MPGs).  This is true both for tensors that admit a non-vanishing component when SOC is turned off (e.g., the altermagnetic texture tensors) and for properties that are purely SOC-driven.  Such angular dependences have been predicted for XMCD in altermagnets \cite{Hariki2024} and are experimentally accessible, since the N\'eel vector can be manipulated by magnetic-field protocols. \cite{Amin2024,Hariki2024} In particular, XMCD experiments on $\alpha$-Fe$_2$O$_3$  \cite{mallon2026revealingoriginxmcdaltermagnet} demonstrated that the weak magnetisation and the  XMCD tensor have  completely different dependences on the direction of the N\'eel vector, despite having the same symmetry properties.  Meanwhile, even in predominantly quasi-collinear magnets, the magnetic order parameter can vary continuously in space, with its direction rotating across domain walls and in magnetic textures.\cite{Cheong2020} The resulting spatial variation of local tensorial properties is becoming increasingly accessible through microscopy at different wavelengths and through scanning-probe techniques.\cite{Cheong2020}

The dependence of SOC-driven properties on the order-parameter direction has motivated several classification schemes of altermagnets \cite{Cheong2024}, which place the same material into different classes depending on the direction of the order parameter.  Much of this discussion has been framed either in terms of magnetic point groups (MPGs) or, more recently, spin groups (SGs), which have been re-introduced in close connection with the concept of altermagnetism.  Ironically, much of this language is rather unfamiliar to the neutron or magnetic X-ray scattering communities, which historically determined all of the magnetic structures that are currently under discussion.  Ever since the 1980s, this community employed the so-called Bertaut--Izyumov framework\cite{Bertaut1968, Bertaut1971, IzyumovNaish1979JMMM,  IzyumovEtAl1979a, IzyumovEtAl1979b, IzyumovEtAl1979c, Izyumov1991}  based on representation theory, in which time-reversal symmetry is not an essential ingredient.  Within the Bertaut--Izyumov framework, magnetic anisotropy is naturally described through exchange multiplets\cite{Izyumov1980}, which comprise the set of irreducible representations activated as the magnetic order parameter is rotated with respect to the crystal lattice.

In this paper, I employ the Bertaut--Izyumov framework and the concept of exchange multiplets to define the angular dependence of generalised time-reversal-odd (TRO) tensors as a function of the direction of the order parameter.  This is achieved through a representation-theoretic framework based on tensorial intertwiners, whose canonical form separates the crystallographic and magnetic aspects of tensor construction.  The resulting canonical intertwiners are determined entirely by representation theory and are independent of the polynomial order of the expansion. Intertwiner-derived tensors provide a continuous description connecting the tensor forms associated with different MPGs, whilst also providing a method to extract the `SOC-free' component, which survives in the absence of magnetic anisotropy, and to classify SOC-related tensor components by the corresponding power of the order parameter.  This framework is equally applicable to both altermagnets and `conventional' zone-centre antiferromagnets, though different time-reversal-odd tensorial properties are allowed in these two classes.  I will provide several examples of tensor construction in quasi-collinear, zone-centre magnets, and also discuss the extension to non-collinear and non-$\Gamma$-point structures.   Finally, I discuss the implications of the present framework for the classification of magnetic structures. I propose a representation-theoretic perspective that naturally connects the spin-group and irreducible-representation languages, connects them naturally to magnetic point groups for specific directions of the order parameters, and removes the false dichotomy between spin groups and magnetic point groups.

\section{The Bertaut--Izyumov method}

In this section, I summarise the Bertaut--Izyumov framework for constructing magnetic structures from \irreps.  For simplicity, I assume a single Wyckoff orbit of magnetic atoms, but the procedure can easily be repeated for multiple orbits.  A Wyckoff orbit of magnetic sites is a set of atomic positions occupied by the same magnetic ion, generated by applying all space-group operations to a single representative site. Each space-group \irrep~is associated with a star of symmetry-equivalent propagation vectors in reciprocal space.   In its simplest and most popular version, the Bertaut--Izyumov method assumes that different propagation vectors $\vec{k}$ within the star correspond to different domains, so that one can limit the discussion to a single $\vec{k}$.  The construction is summarised in Appendix~\ref{app: B-I recipe}. The resulting magnetic structure, i.e. a collection of real magnetic moments associated with every magnetic site $j$ in every unit cell $n$, may be written as:

\begin{equation}
\vec{m}(n,j) =\sum_\alpha c_\alpha \vec{S}^\Delta_\alpha (j) e^{-2 \pi i \vec{k} \cdot \vec{R}_n}+ c.c.
\end{equation} 

where the $\vec{S}^{\Delta}_{\alpha}(j)$ are complex basis vectors associated with magnetic site $j$ and the selected \irrep~$\Delta(\vec{k})$. The index $\alpha$ runs over the dimension and multiplicity of the \irrep, while the coefficients $c_\alpha$ specify the particular magnetic structure within that \irrep.

The Bertaut--Izyumov framework is closely connected to the Landau theory of phase transitions. According to Landau theory, continuous magnetic ordering transitions generally result from the condensation of either a single mode or of multiple modes transforming according to the same \irrep. Consequently, Bertaut--Izyumov magnetic structures provide the natural representation-theoretic description of ordered magnetic phases. The tensorial framework developed below adopts the same philosophy, extending the representation-theoretic description from magnetic structures to the tensorial properties associated with them.

\subsection{Time reversal and magnetic point groups in the Bertaut--Izyumov construction}
\label{sec: MPG_Construction}

As we have seen,  time reversal plays no fundamental role in the Bertaut--Izyumov framework. Indeed, the same construction applies equally well to other order parameters, such as phonon displacement modes.  Time reversal is instead introduced \emph{a posteriori} to determine the magnetic point group (MPG), which is then commonly used to identify the allowed tensorial properties.  The MPG is constructed from the subgroup $H$ of the parent space group $G$ for which the selected \irrep~maps the corresponding \emph{real} magnetic structure onto a one-dimensional real representation.  This subgroup is not, in general, a subgroup of the little group associated with a given propagation vector, because the physical magnetic structure may involve several arms of the propagation-vector star (indeed, it always does whenever $\vec{k}$ and $-\vec{k}$ are inequivalent).  Methods for determining the MPG in such cases have been discussed extensively, particularly in the context of magnetic multiferroics (see, for example, Ref.~\onlinecite{Radaelli2007}).  Once the subgroup $H$ has been identified, the MPG is obtained by retaining only the rotational parts of its symmetry operations and attaching time reversal (a prime) to all operations whose character is $-1$.  If at least one pure translation carries character $-1$, the resulting MPG is paramagnetic (grey), as is always the case for incommensurate magnetic structures.  The MPG construction is generally regarded as the fundamental framework for determining allowed tensorial properties. The remainder of this paper adopts the representation-theoretic viewpoint of Izyumov's exchange multiplets, in which the order parameter is treated as a \emph{continuous} variable rather than as a collection of discrete symmetry directions associated with individual MPGs.

\subsection{Exchange multiplets}
An important extension of the Bertaut--Izyumov framework was introduced by Izyumov in 1980 \cite{Izyumov1980} to address a natural limitation of the conventional Landau picture. Landau theory assumes that different irreducible representations generally have different free energies. But what if this is not the case?  The most obvious counter-example occurs in the absence of any single-site or exchange anisotropy, i.e., in the pure Heisenberg exchange limit.  Here, different magnetic structures related to each other by a global rotation in spin space would have the same magnetic exchange energy, yet, in general, they would belong to different \irreps.  Izyumov presented a recipe to construct all \irreps~belonging to these `exchange multiplets', and also, separately, to construct the magnetic structures themselves.  In essence, exchange multiplets are obtained from a \emph{single irreducible component} of the permutational representation (which we shall call the \emph{generating \irrep}), which is tensor-multiplied with the axial vector representation.  Likewise, the magnetic structures in a multiplet are obtained as tensor products of the permutation space of the generating \irrep~and the axial vector space (see Appendix \ref{Appendix: Perm_rep}).  Crucially, Izyumov showed \cite{Izyumov1980} that, for a given magnetic structure, all structures related to it by a global rotation in spin space belong to the same exchange multiplet.  Note that this construction is completely general, and can be applied to commensurate, incommensurate, collinear and non-collinear magnetic structures alike.   The set of \irreps~in a multiplet forms a true degenerate multiplet only in the absence of magnetic anisotropy, which, in most cases, means in the absence of spin-orbit coupling (SOC).  However, the exchange-multiplet construction itself remains well defined even in the presence of SOC, where it provides a continuous representation-theoretic description of the order parameter despite the lifting of the exact degeneracy. It therefore provides a natural representation-theoretic description of the continuous order-parameter space, and forms the starting point of the tensorial framework developed in the following sections.

\subsection{The case of constant-moment, collinear structures at the zone centre/zone boundary}
\label{sec: constant_moment}

A particularly important case arises when the exchange multiplet describes a constant-moment collinear magnetic structure with propagation vector either at the zone centre ($\Gamma$ point) or at high-symmetry points at the zone boundary.  In these cases, magnetic Wyckoff orbits are \emph{bipartitions} --- in other words, one can distinguish between `up' and `down' sites.  Throughout most of this paper, I will refer to this as `collinear ordering'.  In section \ref{Sec: Landau_Theory_connection}, I will discuss this point further in the context of Landau theory and the coupling with secondary order parameters, and clarify that `collinear' does not refer to the structure as a whole but to the ordering patterns associated with the \emph{primary order parameter}.\newline

The collinearity condition  has a number of important consequences:

\begin{enumerate}
\item The Wyckoff orbits bipartition defines a real, 1D \irrep~(i.e., a sign character $\chi_{bip}(g)$) such that for every element $g$ of the paramagnetic (parent) group $G$, $\chi_{bip}(g)=-1$ if $g$ reverses the direction of the N\'eel vector, while $\chi_{bip}(g)=+1$ otherwise.  
\item If $g$ leaves at least one element of the orbit invariant, then $\chi_{bip}(g)=+1$ necessarily.  Conversely, one can show (see Appendix \ref{Appendix: bicolor_irrep}) that such an \irrep~always induces a bipartition of the Wyckoff orbit into `black' and `white' sites (though the colour label is arbitrary).  Hence, in this context, we shall call this \irrep~the `bipartition \irrep'.
\item Since all magnetic structures in an exchange multiplet share the same bipartition, it naturally follows that the generating \irrep~from which the multiplet is constructed is precisely the bipartition \irrep~$\chi_{bip}$.  Hence, the (reducible) multiplet representation $\Gamma_{mult}$ is the product of the bipartition \irrep~times the axial vector representation $A$ of $G$, i.e.:

\begin{equation}
\label{eq: collinear_multiplet}
\Gamma_{mult}=\chi_{bip} \otimes A
\end{equation}
 
 and all the multiplet \irreps~ are contained in $\Gamma_{mult}$ and are themselves the product of $\chi_{bip}$ times an \irrep~contained in $A$.
 
\item It follows that the general procedure to construct MPGs (discussed in section \ref{sec: MPG_Construction}) is also modified.  Since the permutation component $\chi_{bip}$ is already real and 1D, the MPG is constructed from elements of $G$ that map the axial vector representation into a real 1D representation.  It also follows that zone-boundary magnetic structures of this kind have \emph{paramagnetic} MPGs since for these structures, pure translations have character $-1$.  One often says that these structures retain time reversal symmetry at the MPG level.

\item Finally, it is essential to point out the direct connection between permutation representations and binary spin groups.  Mathematically, the spin group associated with a given exchange multiplet is the graph of the $G \rightarrow \mathbb{Z}_2$ homomorphism created by the bipartition \irrep.  Dropping the group-theoretical language, this means that the spin group is obtained from the paramagnetic group $G$ by composing each operator $g$ having  character $\chi_{bip}=-1$ with a 2-fold rotation around an axis perpendicular to the magnetic moment (wherever it is pointing).

\end{enumerate}

\section{Physical tensors for exchange multiplets}

\subsection{Definition of intertwiners}
Physical tensors (also known as materials tensors) describe either spontaneous ferroic properties (e.g., the electrical polarisation or magnetisation vectors) or physically allowed responses to external fields (e.g., the magnetoelectric tensor).  The general procedure to construct physical tensors is well known, and entails projecting a generic tensor with the expected rank and index symmetry onto the 1D real \irrep~$\chi_{MPG}$ that defines the MPG.  This is equivalent to requiring the tensor to transform according to the totally symmetric representation of the MPG, as required by Neumann's principle.  The connection with exchange multiplets becomes clear when such tensors are expanded in powers of the order parameters (Landau--Lifshitz expansion).  For example, for a rank-2, time-reversal-odd tensor, in the collinear case one may write:

\begin{equation}
T_{ij}=\mathcal{I}^{(1)}_{ijk} L_k + \mathcal{I}^{(3)}_{ijkmn} L_k L_m L_n + \dots
\end{equation}

where  we have chosen the N\'eel vector as the order parameter (this is always possible in the collinear case), and the $\bm{\mathcal{I}}$'s which are tensor-valued coefficients known as \emph{intertwiners}.  The central question addressed in this paper is therefore: what are the symmetry requirements on the tensors $\bm{\mathcal{I}}^{(1)}$, $\bm{\mathcal{I}}^{(3)}$ etc. for $\ten{T}$ to be `physical' for all directions of the order parameter? For constant-moment collinear structures, the answer is remarkably simple and forms the basis of the representation-theoretic framework developed below.

\subsection{Intertwiner theorems} 

\subsubsection{Main intertwiner theorem}

We now determine the symmetry properties of the intertwiners appearing in the Landau--Lifshitz expansion of covariant tensors.  In the Landau--Lifshitz expansion, tensorial properties are expressed as functions of the magnetic order parameter. Invariance of the Landau free energy under every operation of the parent group requires the tensor to transform covariantly with the order parameter, namely

\begin{equation} 
\label{eq: equivariance}
\ten{T}'(\vec{L})=\ten{T}(\vec{L'})
\end{equation}

This condition is the natural tensorial analogue of the invariance of the Landau free energy and is assumed throughout the remainder of this work.  On this basis one can prove the following theorem:

\begin{center}
\fbox{%
\parbox{0.9\linewidth}{%
\textbf{Intertwiner theorem (quasi-collinear structures):} All intertwiners in the Landau--Lifshitz expansion of covariant tensors in powers of the magnetic order parameters transform according to the bipartition \irrep~ $\chi_{bip}$ (for odd powers) or the totally symmetric \irrep~ (for even powers).
}}
\end{center}

The proof, reported in Appendix \ref{Appendix: simple_proofs}, only makes use of the covariance condition and on the fact that even/odd powers of a sign character \irrep~are, respectively,  the totally symmetric \irrep~ and the sign character itself.\newline

\subsubsection{Canonical intertwiners: a corollary to the main intertwiner theorem}

The intertwiner theorem immediately suggests introducing symmetry-adapted representations of the contracted indices.  Therefore, consider an intertwiner $\mathcal{I}$ symmetrised with $\chi_{bip}$ as just discussed, and perform a canonical transformation for a subset of indices onto a symmetry-adapted basis set of those indices, which would typically involve the order parameter and/or one of the physical quantities that contract the intertwiner.  The `canonical intertwiner' will have the form:

\begin{equation}
\left(\mathcal{I}^{(A_{1g})}_{\{i\},1}, \mathcal{I}^{(A_{1g})}_{\{i\},2}, \dots,  \mathcal{I}^{(A_{2g})}_{\{i\},1} \dots\right)
\end{equation}

where $\{i\}$ represents the remaining indices and $1, 2,\dots$ are the different copies of a given \irrep~in the canonical basis.  The following corollary follows immediately from the intertwiner theorem, since a change of basis cannot alter the representation under which the intertwiner transforms:

\begin{center}
\fbox{%
\parbox{0.9\linewidth}{%
\textbf{Corollary ---canonical intertwiners}: Let $\mathcal{I}$ be an intertwiner transforming according to the generating \irrep~$\chi_{bip}$. After expressing any subset of its contracted indices in a symmetry-adapted basis, each canonical block $\mathcal{I}^{(\Gamma)}_{\{i\},\alpha}$ transforms according to the representation $\Gamma \otimes \chi_{bip}$. 
}}
\end{center}

Consequently, the tensor space associated with the remaining indices $\{i\}$ must contain the representation $\Gamma \otimes \chi_{bip}$.  Therefore, standard canonical intertwiner forms can be obtained by projecting that tensor space onto the corresponding isotypic component.\newline

\subsection{Construction of canonical intertwiner forms by `twisted' projection}
\label{sec: twisted projection}

The canonical intertwiner corollary has important practical consequences. Canonical intertwiners may be constructed entirely by group-theoretical means, namely by projecting the tensor space of the remaining indices onto the representation $\Gamma \otimes \chi_{bip}$ (see discussion in Section \ref{sec: twisted projection}). Their form is therefore completely independent of the order of the symmetry-adapted polynomials and on the physical meaning of the tensors. Moreover, different copies of the same \irrep~possess identical canonical forms; only the number  of repeated forms (each with independent coefficients) changes, being equal to the multiplicity of $\Gamma \otimes \chi_{bip}$ in the tensor representation of the remaining indices. All crystallographic complexity is thus transferred to the symmetry-adapted polynomial basis, which depends only on the parent point group and is completely unrelated to magnetic ordering.

When constructing canonical intertwiners, we require them not only to transform with the $\Gamma \otimes \chi_{bip}$ representation, but also to transform with the same matrices as the corresponding canonical polynomials.  Let these matrices for a given parent group element $g$ and  \irrep~$\Gamma$ be $D^{\Gamma}_{\mu\nu}(g)$.  Then, the canonical projection operators are:

\begin{equation}
\label{eq: twisted projector}
\bm{\mathcal{P}}^{\Gamma}_{\mu\nu}=\frac{d^{\Gamma}}{h_G}\sum_g \left[\chi_{bip}(g)\,D^{\Gamma}_{\mu\nu}(g)\right]^* \ten{R}(g)
\end{equation}

where $d^{\Gamma}$ is the dimension of $\Gamma$, $h_G$ is the order (number of elements) of the parent group $G$, and $\ten{R}(g)$ is the usual generalised tensor rotation operator.  One may also think of Equation \ref{eq: twisted projector} as an ordinary projector over a `twisted' representation with matrices $\chi_{bip}(g)\,D^{\Gamma}_{\mu\nu}(g)$, where the twisting is provided by the bipartition sign character $\chi_{bip}$.  

\subsection{Centrosymmetric groups}

In centrosymmetric groups, \irreps\, canonical forms and  canonical polynomials have distinct parities $\pi$, either \emph{gerade} or \emph{ungerade}.  One can immediately see that the projector in Equation \ref{eq: twisted projector} yields zero unless $\pi(\chi_{bip}) \times \pi_F \times \pi(\Gamma) =1$, where $\pi_F$ is opposite for polar and axial tensors of the same rank.  Moreover, for a \emph{gerade} polynomial (i.e., all even-order polynomials and odd-order polynomial of an axial vector such as the N\'eel vector), only $\Gamma_g$'s will be active, while for a \emph{ungerade} polynomial (i.e., odd-order polynomial of a polar vector) only $\Gamma_u$'s will be active.  Since the projected form is the same regardless of the individual parities, in the remainder the projected forms will be labelled $g/u$, for example, $M^{(E_{g/u})}$.

\subsection{The isotropic case: co-rotating tensors and absolute intertwiners}

\subsubsection{Uncontracted forms and absolute intertwiners}

As we have just seen, any covariant tensor can be expanded in powers of the order parameter, and any term in the expansion is the contraction of the appropriate polyadic form in $\vec{L}$ (e.g., $L_k L_m L_n$) with an intertwiner tensor $\bm{\mathcal{I}}$.   In the collinear case, the order parameter can always be taken as the N\'eel vector and the tensor  $\bm{\mathcal{I}}$ transforms as either the bipartition \irrep~(odd powers) or the totally symmetric \irrep~(even powers).  

In the absence of magnetic anisotropy, the order parameter can only be contracted through fully isotropic tensors, i.e. products of  $\delta_{ij}$, $\epsilon_{ijk}$, etc.\cite{Spencer1971} to yield the spin indices.  The remaining tensor indices, however, need not be isotropic and may retain the full point-group dependence of the crystal. Thus, the order-parameter dependence takes the uncontracted forms illustrated below.

\begin{eqnarray}
\label{eq: example_co-rotating}
\beta^{(1)}_{ij}&=& \delta_{ik}\mathcal{A}_j L_k\nonumber\\
\gamma^{(1)}_{ijk}&=& \epsilon_{ijl}\mathcal{A}_k L_l\nonumber\\
\gamma^{(2)}_{ijk}&=& \delta_{il}\mathcal{A}_{jk} L_l\nonumber\\
&&\dots
\end{eqnarray}

The tensor $\bm{\mathcal{A}}$ appearing in Equation \ref{eq: example_co-rotating} can be called an `absolute' (unbound) intertwiner, in the sense that it is not bound to (contracted with) the order parameter.  In this list, one may recognise the magneto-electric tensor $\beta^{(1)}$ and the piezomagnetic or altermagnetic tensor (d-wave) $\gamma^{(2)}$ (the $j$ and $k$ indices must be symmetrised), while $\gamma^{(1)}_{ijk}$ can represent a non-linear optics tensor.  It is also clear by inspection that the corresponding intertwiners are 
$ \delta_{ik}\mathcal{A}_j $, $  \epsilon_{ijl}\mathcal{A}_k$, etc.
\newline

By contrast, the expression for the spontaneous Faraday (or MOKE) antisymmetric tensor:

\begin{equation}
\beta^{(2)}_{ij}= \left(\delta_{ik}\delta_{jm}-\delta_{im}\delta_{jk}\right) \mathcal{A}_m L_k\nonumber\\
\end{equation}

is \emph{not} allowed in the isotropic case.  In fact, while the first part :

\begin{equation}
\delta_{ik}\delta_{jm} \mathcal{A}_m L_k=L_i A_j
\end{equation}

is co-rotating (we will encounter this very expression in the SOC-free component of the magneto-electric tensor), the second part:

\begin{equation}
-\delta_{im}\delta_{jk} \mathcal{A}_m L_k= \mathcal{A}_i L_j
\end{equation}

has the spin index $i$ determined by the vector $\mathcal{A}_m$ and unrelated to the direction of $\vec{L}$.

From the intertwiner theorem, one can easily derive the following (see Appendix \ref{Appendix: simple_proofs}):

\begin{enumerate}
\item The resulting intertwiners are covariant iff the tensors $\bm{\mathcal{A}}$ transform as the bipartition \irrep.
\item These tensors are not only covariant but also  \emph{co-rotating} with $\vec{L}$, i.e.,a generic tensor of this form

\begin{equation}
\bm{\mathcal{A}}(\mat{R} \vec{L}) = D(\mat{R}) \bm{\mathcal{A}}(\vec{L}) 
\end{equation}

where $\mat{R}$ is a rotation matrix and $D(\mat{R})$ is a representation of $\mat{R}$ appropriate for the rank of the `spin indices' of the resulting tensor $\ten{T}$.  

For example, for two of the tensors in eq. \ref{eq: example_co-rotating}, 

\end{enumerate}

\begin{equation}
 \delta_{ik} \mathcal{A}_{j} \left(R_{k k'} L_{k'}\right) = \mathcal{A}_{j} R_{i k'} L_{k'}= \mathcal{A}_{j} R_{i k'} \delta_{k'k''} L_{k''} = R_{i i'} \left( \delta_{i' k}  \mathcal{A}_j\right) L_k
 \end{equation}
 
 \begin{equation}
 \epsilon_{ijl}  \mathcal{A}_{j} \left(R_{l l'} L_{l'}\right) =  \det{(R)} R_{i i'} R_{j j'} \left(\epsilon_{i'j'l}  \mathcal{A}_{j} \right) L_l
 \end{equation}

 More generally, one may write:

\begin{equation}
\ten{T}=\ten{X} (\vec{L}) \otimes \bm{\mathcal{A}}
\end{equation}

where $\ten{X} (\vec{L})$ is an odd-rank tensor polynomial in $\vec{L}$, such as $L_i$, $L_i L_j L_k$ etc., but also $\epsilon_{ijk}L_k$ etc., provided that none of these indices are contracted with $\bm{\mathcal{A}}$.  

Although co-rotating higher-order tensors with multiple spin indices exist in principle, here no example will be shown.  However, co-rotating tensors of the following form are always allowed:

\begin{eqnarray}
\label{eq: example_co-rotating_higherL}
\beta_{ij}&= \delta_{ik} \mathcal{A}^{(1)}_j L_k +  \delta_{ik} \delta_{lm} \mathcal{A}^{(3)}_j L_k L_L L_m+  \delta_{ik} \delta_{lm} \delta_{np} \mathcal{A}^{(5)}_j L_k L_L L_m L_n L_p +\dots\nonumber\\
&=\delta_{ik} \left(\mathcal{A}^{(1)}_j +|L|^2 \mathcal{A}^{(3)}_j + \mathcal{A}^{(5)}_j |L|^4+\dots \right)L_k 
\end{eqnarray}
 
Since the form of the $\mathcal{A}^{(i)}$ is always the same, this is equivalent to saying that the \emph{coefficients} of the $\mathcal{A}$ tensors can themselves be expanded in even powers of the modulus of the order parameter.
 
\subsubsection{Extracting co-rotating components}

When a co-rotating component is allowed, it is always possible to determine its tensorial form without performing the whole Landau-Lifshitz tensor expansion.  For example, in the case of the tensor in eq. \ref{eq: rank2_expansion}, one may write the expansion as:

\begin{equation}
\label{eq: rank2_expansion_cor}
\alpha_{ij}(\vec{L})= \mathcal{I}^{(1)}_{ijk} L_k +  \mathcal{I}^{(3)}_{ijklm} L_kL_m L_n + \dots = \left(\delta_{ik} \mathcal{A}_j+\tilde{ \mathcal{I}}^{(1)}_{ijk}\right) L_k +  \mathcal{I}^{(3)}_{ijklm} L_kL_m L_n + \dots 
\end{equation}

where the first term in bracket is the co-rotating part. To construct this, it is sufficient to symmetrise $\bm{ \mathcal{A}}$ in the usual way with the bipartition \irrep. 

When appropriate, we can also make the notation more specific to indicate not only the rank but also the Jahn symbol\cite{Jahn1949} of the intertwiner.  For example, for the spin texture intertwiners we may substitute:

\begin{equation}
s_i (\vec{k}, \vec{L})= \left( \mathcal{I}^{(1[2]1)}_{ijkl}  \,k_jk_k +\mathcal{I}^{(1[4][1])}_{ijkmnl}  \,k_jk_k k_mk_n + \dots \right) L_l
\end{equation}

The co-rotating component is at the first order in  $\vec{L}$, and there is no need to expand further in $\vec{L}$:

\begin{equation}
s^{alt}_i(\vec{k}, \vec{L})=L_i \left( \mathcal{A}^{([2])}_{jk}k_jk_k +\mathcal{A}^{([4])}_{jklm}k_jk_kk_lk_n + \dots \right)
\end{equation}


where, once again, the tensors $\bm{\mathcal{A}}$'s are straightforwardly obtained by symmetrisation with the bipartition \irrep.

It is important to emphasise that \emph{a co-rotating component  does not always exist}.  For example, in certain symmetries the weak magnetisation can be written as:

\begin{equation}
m_i=\epsilon_{ijk}L_jD_k
\end{equation}

where $\vec{D}$ is a macroscopic Dzyaloshinskii-Morija vector, and the corresponding spontaneous Faraday rotation tensor (which is relevant for both MOKE and XMCD) as:

\begin{equation}
\chi_{ij}=(\delta_{ik}\delta_{jm}-\delta_{im}\delta_{jk}) L_jD_k
\end{equation}

In both cases, $\vec{L}$ is irreducibly contracted with $\vec{D}$ and no co-rotating component exists.

\subsubsection{Absolute canonical intertwiners}
\label{sec: canonical_absolute}

As discussed above, intertwiners in fully canonical form reduce to covariant scalars  $\sigma^{(\Gamma)}$, and the canonical intertwiner theorem implies that only the channel with  $\Gamma=\chi_{bip}$ survives (Equation \ref{eq: condition_covariant_scalars}).  Since absolute intertwiners carry neither order-parameter nor spin indices, their canonical representation on the physical variables is already fully reduced. Consequently, only the symmetry-adapted polynomial components transforming according to $\chi_{bip}$ contribute to absolute intertwiners. This leads to remarkably simple expressions:  for example, in the case of spin textures, one can write the `altermagnetic' component as:

\begin{equation}
\label{eq: Canonical_scalars}
s_i^{alt}(\vec{k})=\delta_{ij} L_j \sum_{\alpha} \sigma^{(\chi_{bip})}_\alpha \mathcal{P}^{(\chi_{bip})}_\alpha(\vec{k})
\end{equation}

where $\sigma^{(\chi_{bip})}_\alpha$ are covariant scalar coefficients and $ \mathcal{P}^{(\chi_{bip})}_\alpha(\vec{k})$ are polynomial in $\vec{k}$ all transforming with the same \irrep~$\chi_{bip}$.

\section{Representation-theoretic constructions of tensor properties}
\subsection{Construction of the Landau--Lifshitz expansions}

This is the most obvious and perhaps most useful application of the canonical intertwiners framework. While Cartesian intertwiners quickly become unmanageable as the order in $\vec{L}$ grows, the corresponding canonical forms retain the familiar Cartesian tensor representation of the physical tensors. The only difference is that their coefficients become symmetry-constrained functions of the order parameter.  At a given order in $\vec{L}$, one writes:

\begin{equation}
\ten{T}=\left(\mathcal{I}^{(A_{1g})}_{\{i\},1}, \mathcal{I}^{(A_{1g})}_{\{i\},2}, \dots,  \mathcal{I}^{(A_{2g})}_{\{i\},1} \dots\right)\left(\mathcal{P}^{(A_{1g})}_{1}(\vec{L}), \mathcal{P}^{(A_{1g})}_{2}(\vec{L}),\dots, \mathcal{P}^{(A_{2g})}_{1}(\vec{L}), \dots\right)^{\mathrm T}
\end{equation} 

where $\mathcal{I}^{(\Gamma)}_{\{i\},\alpha}$ is the canonical intertwiner form for copy $\alpha$ of \irrep~$\Gamma$, and the $ \mathcal{P}^{(\Gamma)}_\alpha(\vec{L})$ are symmetry-adapted polynomials in the order parameter.

For sufficiently high order (3\textsuperscript{rd} order for most point groups, 5\textsuperscript{th} order for the cubic point groups), the polynomial expansion in $\vec{L}$ contains all irreducible representations of the parent point group.  In this case, the Cartesian tensor $T\left(\boldsymbol{L}\right)$ is the most general tensor field compatible with the parent symmetry. Evaluation at any particular order-parameter direction automatically yields the same tensor form that would be obtained independently from the corresponding magnetic point group.  MPG selection rules are recovered automatically by the nodes of those polynomials combined with the canonical forms.

\subsection{Construction of the co-rotating component and the link with altermagnetism}

\subsubsection{Cartesian forms of the absolute intertwiners}

Since the linear polynomials $L_x, L_y, L_z$ are both Cartesian and canonical (though of course not always single-\irrep\ as a set),  one can extend the discussion of the previous paragraph to the construction of co-rotating forms.  Absolute intertwiners are one rank lower than the physical tensors and two ranks lower than the linear Cartesian intertwiners, and are therefore extremely easy to construct.  The co-rotating physical tensors are reconstructed from expressions such as those in Equation \ref{eq: example_co-rotating}.  For example, the  physical tensor is obtained from 

\begin{equation}
\delta_{ik}\mathcal{A}_j L_k
\end{equation}

by replicating $\mathcal{A}$ in three identical row blocks and multiplying each block by $L_x, L_y, L_z$, respectively.

\subsubsection{Canonical transformation of the physical variables}

Although the co-rotating component of a tensor is most easily obtained as discussed in the previous paragraph, this method does not allow one to separate cleanly this component from the rest of the tensor.  This task is most easily accomplished by employing a canonical basis of the physical quantities, since the intertwiner theorem does not distinguish between those and the order parameter.

One may write:

\begin{eqnarray}
\label{eq: rank2_expansion_canonical}
s_i(\vec{L}, \vec{k})&=&\mathcal{J}_i \cdot \left(\mathcal{P}^{(A_{1g})}_{1}(\vec{k}), \mathcal{P}^{(A_{1g})}_{2}(\vec{k}),\dots, \mathcal{P}^{(A_{2g})}_{1}(\vec{k}), \dots\right)^{\mathrm T}\nonumber\\
\mathcal{J}_i&=&\left(\mathcal{I}^{(1) (A_{1g})}_{ik,1}, \mathcal{I}^{(1)(A_{1g})}_{ik,2}, \dots,  \mathcal{I}^{(1) (A_{2g})}_{ik,1} \dots\right) L_k + \left(\mathcal{I}^{(3) (A_{1g})}_{1,klm}, \mathcal{I}^{(3)(A_{1g})}_{2,klm}, \dots,  \mathcal{I}^{(3) (A_{2g})}_{1,klm} \dots\right)L_kL_l L_m + \dots
\end{eqnarray}

where the canonical basis is employed for the physical variables to be contracted (in this example case, the wavevector $\vec{k}$) rather than the order parameters.  This construction cleanly separates the co-rotating (SOC-free) from the non-corotating tensors at the linear order in $\vec{L}$.  In fact, as we have seen, the only intertwiner allowed in the absence of anisotropy is $\mathcal{I}^{(1)(\chi_{bip})}_{ik,r}$, where $r$ denotes (potentially) multiple copies of the \irrep~$\chi_{bip}$ in the intertwiner decomposition.  Equation \ref{eq: rank2_expansion_canonical} mirrors Equation \ref{eq: Canonical_scalars}, so $\mathcal{I}^{(1)(\chi_{bip})}_{ik,r}$ is a canonical scalar and contains the SO(3)- invariant term $\tau \delta_{ik}$, where $\tau = Tr\left(  \mathcal{I}^{(1)(\chi_{bip})}_{ik,r}\right)$.  We can therefore write:

\begin{equation}
\mathcal{I}^{(1)(\chi_{bip})}_{ik,r} =\tau \delta_{ik}+\tilde{\mathcal{I}}^{(1)(\chi_{bip})}_{ik,r}
\end{equation}

where $\tilde{\mathcal{I}}^{(1)(\chi_{bip})}_{ik,r}$ is the `residual' traceless part, which transforms with the same \irrep\ but is \emph{not} co-rotating.  This procedure can be extended to higher orders in $\vec{L}$, since the series:

\begin{equation}
\left(\tau_1+\tau_2 |L|^2+ \tau_3 |L|^4+\dots\right)  \delta_{ik} 
\end{equation}

represents the expansion in powers of $|L|^2$ of the co-rotating tensor on this canonical basis.

\subsection{Intertwiners on other canonical bases}

\subsubsection{Vector intertwiners on a mixed canonical basis}
\label{Sec: Mixed Canonical Bases}

One may also employ mixed canonical bases of the order parameter and the physical variables, leaving only spin indices in Cartesian form:

\begin{equation}
s_i(\vec{L}, \vec{k})=\left(\mathcal{I}^{(A_{1g})}_{i,1}, \mathcal{I}^{(A_{1g})}_{i,2}, \dots,  \mathcal{I}^{(A_{2g})}_{i,1} \dots\right)\left(\mathcal{P}^{(A_{1g})}_{1}(\vec{L},\vec{k}), \mathcal{P}^{(A_{1g})}_{2}(\vec{L},\vec{k}),\dots, \mathcal{P}^{(A_{2g})}_{1}(\vec{L},\vec{k}), \dots\right)^{\mathrm T}
\end{equation}

These intertwiners are the \emph{canonical vectors} discussed below.  Their forms are particularly simple and can be used as a computationally efficient way to construct very complex tensors.  In this context, a useful development is to construct canonical vectors on dyadic (or polyadic) products of symmetry-adapted polynomial bases, each in a \emph{single} variable.  This can be done by a straightforward application of the point-group Clebsch-Gordan coefficients, as demonstrated in Appendix \ref{App: Fully_Canonical_CG}.

\subsubsection{Scalar intertwiners on a fully canonical basis}

Finally, all contracted variables, including the spin indices, may be expressed in symmetry-adapted form. The intertwiner then carries no remaining Cartesian tensor indices and reduces to a collection of \emph{covariant scalars}:

\begin{equation}
\ten{T}
=
\left(
\sigma^{(A_{1g})}_{1},
\sigma^{(A_{1g})}_{2},
\dots,
\sigma^{(A_{2g})}_{1},
\dots
\right)
\left(
\mathcal{P}^{(A_{1g})}_{1},
\mathcal{P}^{(A_{1g})}_{2},
\dots,
\mathcal{P}^{(A_{2g})}_{1},
\dots
\right)^{\mathrm T},
\end{equation}

where the $\sigma^{(\Gamma)}_{\alpha}$ are the fully reduced canonical intertwiners. As shown below, these quantities are not invariants but covariant scalars transforming according to the generating \irrep~$\chi_{\mathrm{bip}}$. Since a fully symmetry-adapted scalar $\sigma^{(\chi_{\Gamma})}$ belongs to the totally symmetric representation $A_ {1g}$, the intertwiner theorem (see next paragraph) imposes that corresponding canonical block must satisfy:
 
\begin{equation}
\chi_{bip} \otimes \Gamma=A_{1g}
\end{equation}

For quasi-collinear magnetic structures, where $\chi_{bip}$  is one-dimensional and real, this immediately implies

\begin{equation}
\label{eq: condition_covariant_scalars}
 \Gamma=\chi_{bip}
\end{equation}

This fully canonical decomposition is closely analogous to the Wigner--Eckart factorization of spherical tensors. In both constructions, the symmetry dependence is carried entirely by Clebsch--Gordan coefficients, while the remaining quantities are reduced objects. In the present case, these reduced objects are the covariant scalars $\sigma^{(\chi_{bip})}$ , which replace the reduced matrix elements of the Wigner--Eckart theorem.

\section{Computational implementation}

In the remainder of this paper, I will construct intertwiners for the parent group $\hmn{-3m}$ and its two non-trivial 1D \irreps~ ($A_{2g}$ and $A_{2u}$), both in fully Cartesian form and with different canonical representations.  For this, I have employed the following computational approach:

\begin{description}
\item[Fully Cartesian intertwiners] these have all been constructed using the program MTENSOR in the Bilbao crystallographic server \cite{PerezMato2015, Gallego2019}.  Although MTENSOR is designed to produce fully symmetric tensors for individual MPGs, it can easily be adapted to  symmetrise tensors with bipartition \irreps~of any crystallographic parent group.  This adaptation is explained in Appendix \ref{App: MTENSOR_Adaptation}.  Using the Bilbao conventions ensures both consistency of notation and reproducibility.

\item[Canonical forms] these have been produced by the twisted projector method (Equation \ref{eq: twisted projector}).  This has been implemented in a completely generic form using a piece of code written in Mathematica \cite{Mathematica}, which is available as part of the Supplementary Information \cite{MyPaperSI}.

\item[Canonical intertwiners] these have been obtained from the fully Cartesian (Bilbao) intertwiners by a coordinate transformation to the appropriate canonical basis. The coordinate transformation has also been implemented in a completely generic form using a second piece of code written in Mathematica \cite{Mathematica}, which is available as part of the Supplementary Information \cite{MyPaperSI}.  I have then verified that the canonical intertwiners associated with a given canonical polynomial correspond to the canonical form for the relevant \irrep, and provided a correspondence table between the parameters of the Cartesian intertwiners (from Bilbao) and those of the canonical forms (from my Mathematica code).  In the most complex cases, I have also verified explicitly that the canonical intertwiners project correctly onto themselves when Equation \ref{eq: twisted projector} is applied.  A further discussion about a computational implementation of canonical vectors through Clebsch-Gordan coefficients is provided in Appendix \ref{App: Fully_Canonical_CG}.

\item[Symmetry-adapted polynomials] The polynomials employed in this work (up to
4\textsuperscript{th} order) are relatively simple and were initially constructed
by hand. In the presence of \irreps\ occurring more than once, the choice of
basis within a given \irrep\ sector is not unique. A rational choice should,
however, preserve the radial sector (e.g., $(|L|^{2n}L_x,|L|^{2n}L_y)$ for an
$E$ \irrep) and retain recognizable polynomial families of the form
$Q L_x,Q L_y$, where $Q$ is a group invariant. An iterative constructor can be
built using the Fischer (or apolar) inner product of homogeneous polynomials:
\begin{equation}
\langle P,Q\rangle_F
=
P\!\left(
\frac{\partial}{\partial L_x},
\frac{\partial}{\partial L_y},
\frac{\partial}{\partial L_z}
\right)
Q(L_x,L_y,L_z)\bigg|_{\mathbf L=0},
\end{equation}
which provides a rotation-invariant inner product on polynomial space
\cite{Render2008}.  An example Mathematica code is provided as part of the Supplementary Information \cite{MyPaperSI}.

\end{description}

\subsection{Computational implementation of the Izyumov formalism}

A Mathematica application has been developed to construct fully covariant vectors, $3 \times 3$ tensors and $3 \times 6$ Voigt tensors (both polar and axial) for all 32 crystallographic point groups and their one-dimensional \irreps~generating Wyckoff-site bipartitions. Crystallographically distinct settings of the point groups, such as 321 and 312, are explicitly included. For a chosen point group and one-dimensional \irrep, the user can select the tensor type and the order of the Landau--Lifshitz expansion, up to fifth order. The resulting tensors can be obtained either as angle-dependent expressions, as functions of the direction of the order parameter, or as fixed tensors for specified order-parameter directions.
The application also provides the corresponding space group and magnetic point group, together with a converter to the new crystallographic convention for magnetic point groups, allowing direct comparison with the results of MTENSOR. Covariance has been explicitly tested in all cases, and consistency with MTENSOR has been verified for many directions of the order parameter. The code has been deployed on Wolfram Cloud and is being prepared for public release. A beta version is available upon request.  For polar and axial vectors, such as weak magnetisation and induced polarisation, the Mathematica notebook also provides a spherical visualisation of the angle-dependent vectors, with colour coding to indicate the symmetry. This visualisation is currently available in the notebook version but has not yet been implemented in the Wolfram Cloud deployment.

\section{Canonical polynomials for the point group $\bar{3}m$} 
\label{Sec: basis transformation}

As we have just seen, time-reversal odd tensors in antiferromagnets can be written as expansion in powers of the order parameter $\vec{L}$,  
\begin{equation}
\ten{T}(\vec{L})= \bm{\mathcal{I}}^{(1)}_i L_i+\bm{\mathcal{I}}^{(3)}_{i,j,k} L_i L_j L_k +\dots
\end{equation}

the appropriate intertwiners $\bm{\mathcal{I}}^{(n)}_{ijk\dots} $ having been symmetrised with the bipartition \irrep.  We have also seen that it is very convenient to write the intertwiners in canonical form, so that the tensors are written as linear combinations of symmetry-adapted basis functions.  Since in the remainder we will discuss examples with parent  point-group symmetry \hmn{-3m}, in this section I present a selection of canonical polynomial bases for this group. In the examples below, care was taken to define the polynomial basis functions such that the matrix transformations for the $E_g$ representations are all the same and are consistent with the transformation matrices of the $(L_x,L_y)$ pair. 

\subsubsection{Canonical polynomials in the order parameter}

For the cubic term in the order parameter expansion, one may use cubic polynomials in $\vec{L}$ as the canonical variables, and write a generic tensor as: 

\begin{equation}
\ten{T}^{(3)}=
M^{(A_{1u})}\Phi^{(A_{1u})}
+\sum_{r=1}^{3}
M^{(A_{2u})}_{r}\,
\Phi^{(A_{2u})}_{r}
+\sum_{r=1}^{3}
\sum_{\alpha=1}^{2}
M^{(E_u)}_{r,(\alpha)}\,
\Phi^{E_u}_{r,(\alpha)}.
\end{equation}

where $r$ runs over the different copies of a given \irrep, $\alpha$ runs over the dimensions of the \irrep~and the ten functions $\Phi^{(\Gamma)}_{n,(\alpha)}$ have been chosen as one of the possible basis sets for the \irreps~of the parent group that span cubic polynomials:

\begin{align}
\label{Eq: l_cube_poly}
\Phi^{(A_{1g})}_1 &= L_x\!\left(L_x^2 - 3L_y^2\right)\nonumber\\
\Phi^{(A_{2g})}_1 &= L_y\!\left(3L_x^2 - L_y^2\right)\nonumber\\
\Phi^{(A_{2g})}_2 &= L_z\!\left(L_x^2 + L_y^2 + L_z^2\right)\nonumber\\
\Phi^{(A_{2g})}_3 &= L_z\!\left(L_x^2 + L_y^2 - L_z^2\right)\nonumber\\
\Phi^{(E_{g})}_{1,(1)} &= L_x\!\left(L_x^2 + L_y^2 + L_z^2\right),
&\qquad
\Phi^{(E_{g})}_{1,(2)} &= L_y\!\left(L_x^2 + L_y^2 + L_z^2\right)
\nonumber\\
\Phi^{(E_{g})}_{2,(1)} &= L_x\!\left(L_x^2 + L_y^2 - L_z^2\right),
&\qquad
\Phi^{(E_{g})}_{2,(2)} &= L_y\!\left(L_x^2 + L_y^2 - L_z^2\right)
\nonumber\\
\Phi^{(E_{g})}_{3,(1)} &= 2L_xL_yL_z,
&\qquad
\Phi^{(E_{g})}_{3,(2)} &= L_z\!\left(L_x^2 - L_y^2\right)
\end{align}

One can also observe that:

\begin{equation}
\left(\Phi^{(E_{g})}_{1,(1)} , \Phi^{(E_{g})}_{1,(2)}, \Phi^{(A_{2g})}_2 \right)=L^2 \left(L_x, L_y, L_z\right)^{\mathrm T}
\end{equation}

Hence, terms in $\ten{T}^{(3)}$ containing $\Phi^{(E_{g})}_{1,(1)} , \Phi^{(E_{g})}_{1,(2)},  \Phi^{(A_{2g})}_2 $ are formally identical to the lower-order $\ten{T}^{(1)}$ multiplied by $L^2$, though the matrix coefficients will be different at the two orders for a given material.  \newline

\subsubsection{Canonical polynomials in the wavevector}

When dealing specifically with spin textures or other $\vec{k}$-dependent properties, one may likewise use polynomials in powers of $\vec{k}$ as the canonical variables.  For this, we define to second and fourth order:

\begin{align}
\label{eq: Xi_poly}
\Xi^{(A_{1g})}_1
&=
k_x^2+k_y^2+k_z^2
\nonumber\\
\Xi^{(A_{1g})}_2
&=
2k_z^2-k_x^2-k_y^2
\nonumber\\[1ex]
\Xi^{(E_g)}_{1,(1)}
&=k_x^2-k_y^2,
&\qquad
\Xi^{(E_g)}_{1,(2)}
&=-2k_xk_y
\nonumber\\
\Xi^{(E_g)}_{2,(1)}
&=
k_yk_z,
&\qquad
\Xi^{(E_g)}_{2,(2)}
&=
-k_xk_z
\end{align}

and

\begin{align}
\label{eq: Psi_poly}
\Psi^{(A_{1g})}_1
&=
\left(k_x^2+k_y^2+k_z^2\right)^2
\nonumber\\
\Psi^{(A_{1g})}_2
&=
\left(k_x^2+k_y^2+k_z^2\right)\left(2k_z^2-k_x^2-k_y^2\right)
\nonumber\\
\Psi^{(A_{1g})}_3
&=
3k_z^4
-6k_z^2\!\left(k_x^2+k_y^2\right)
+\left(k_x^2+k_y^2\right)^2
\nonumber\\
\Psi^{(A_{1g})}_4
&=
k_z\!\left(3k_x^2k_y-k_y^3\right)
\nonumber\\[1ex]
\Psi^{(A_{2g})}_1
&=k_z\!\left(k_x^3-3k_xk_y^2\right)
\nonumber\\[1ex]
\Psi^{(E_g)}_{1,(1)}
&=
\left(k_x^2+k_y^2+k_z^2\right)\left(k_x^2-k_y^2\right),
&\qquad
\Psi^{(E_g)}_{1,(2)}
&=
-2\left(k_x^2+k_y^2+k_z^2\right)k_xk_y
\nonumber\\
\Psi^{(E_g)}_{2,(1)}
&=
\left(k_x^2+k_y^2+k_z^2\right)k_yk_z,
&\qquad
\Psi^{(E_g)}_{2,(2)}
&=
-\left(k_x^2+k_y^2+k_z^2\right)k_xk_z
\nonumber\\
\Psi^{(E_g)}_{3,(1)}
&=
\left(2k_z^2-k_x^2-k_y^2\right)\left(k_x^2-k_y^2\right),
&\qquad
\Psi^{(E_g)}_{3,(2)}
&=
-2\left(2k_z^2-k_x^2-k_y^2\right)k_xk_y
\nonumber\\
\Psi^{(E_g)}_{4,(1)}
&=
\left(2k_z^2-k_x^2-k_y^2\right)k_yk_z,
&\qquad
\Psi^{(E_g)}_{4,(2)}
&=
-\left(2k_z^2-k_x^2-k_y^2\right)k_xk_z
\nonumber\\
\Psi^{(E_g)}_{5,(1)}
&=
k_x^4-6k_x^2k_y^2+k_y^4,
&\qquad
\Psi^{(E_g)}_{5,(2)}
&=
4\!\left(k_x^2-k_y^2\right)k_xk_y.
\end{align}

\subsubsection{Mixed-variable canonical polynomials}
\label{sec: mixed_var}

Finally, one may employ mixed-variable canonical polynomials such as:

\begin{alignat}{2}
\label{eq: Clebsch_Gordan}
Z_1^{(A_{1g})}
&=
L_x \Xi_{s,(1)}^{(E_{g})}
+
L_y \Xi_{s,(2)}^{(E_{g})}
&\nonumber\\[1ex]
Z_1^{(A_{2g})}
&=
L_z \Xi_s^{(A_{1g})}
&\nonumber\\
Z_2^{(A_{2g})}
&=
L_x \Xi_{s,(2)}^{(E_{g})}
-
L_y \Xi_{s,(1)}^{(E_{g})}
&\nonumber\\[1ex]
Z_{1,(1)}^{(E_g)} &= L_x\Xi_s^{(A_{1g})},
&\qquad
Z_{1,(2)}^{(E_g)} &= L_y\Xi_s^{(A_{1g})},
\nonumber\\
Z_{2,(1)}^{(E_g)} &= L_z \Xi^{(E_g)}_{s,(2)},
&\qquad
Z_{2,(2)}^{(E_g)} &= -L_z \Xi^{(E_g)}_{s,(1)}\nonumber\\
Z_{3,(1)}^{(E_g)} &= L_x\Xi^{(E_g)}_{s,(1)}-L_y\Xi^{(E_g)}_{s,(2)},
&\quad
Z_{3,(2)}^{(E_g)} &=- L_x\Xi^{(E_g)}_{s,(2)}-L_y\Xi^{(E_g)}_{s,(1)}
\end{alignat}

where we omitted the repetition of multiple \irrep~copies $s$ of the $\Xi^{(\Gamma)}_s$ polynomials.  Taking the $s$ multiplicity into account, Equation \ref{eq: Clebsch_Gordan} comprises 18 functions, ($2 A_{1g} \oplus 4 A_{2g} \oplus 6 E_g )$,  which is exactly 3 for  ($L_x, L_y, L_z)$ times  6 polynomials in Equation \ref{eq: Xi_poly}.  This expression enables the tensor representation of canonical vectors to be obtained (see Section \ref{Sec: Canonical_vectors_3m}). \newline

More generally, when polynomials in $\vec{L}$ and $\vec{k}$ contain other \irreps, one writes mixed polynomials such as:

\begin{alignat}{2}
\label{eq: Clebsch_Gordan1b}
H_1^{(A_{1g})}
&=
\Phi^{(A_{1g})}_{r} \Xi_s^{(A_{1g})}
&\nonumber\\
H_2^{(A_{1g})}
&=
\Phi^{(E_{g})}_{r,(1)} \Xi_{s,(1)}^{(E_{g})}
+
\Phi^{(E_{g})}_{r,(2)} \Xi_{s, 2}^{(E_{g})}
&\nonumber\\[1ex]
H_1^{(A_{2g})}
&=
\Phi^{(A_{2g})}_{r} \Xi_s^{(A_{1g})}
&\nonumber\\
H_2^{(A_{2g})}
&=
\Phi^{(E_{g})}_{r,(1)} \Xi_{s,(2)}^{(E_{g})}
-
\Phi^{(E_{g})}_{r,(2)} \Xi_{s,(1)}^{(E_{g})}
&\nonumber\\[1ex]
H_{1,(1)}^{(E_g)} &= \Phi^{(E_{g})}_{r,(1)}\Xi_s^{(A_{1g})},
&\qquad
H_{1,(2)}^{(E_g)} &= \Phi^{(E_{g})}_{r,(2)}\Xi_s^{(A_{1g})},
\nonumber\\
H_{2,(1)}^{(E_g)} &= \Phi^{(A_{1g})}_{r} \Xi^{(E_g)}_{s,(1)},
&\qquad
H_{2,(2)}^{(E_g)} &= \Phi^{(A_{1g})}_{r} \Xi^{(E_g)}_{s,(2)}\nonumber\\
H_{3,(1)}^{(E_g)} &= \Phi^{(A_{2g})}_{r} \Xi^{(E_g)}_{s,(2)},
&\qquad
H_{3,(2)}^{(E_g)} &= -\Phi^{(A_{2g})}_{r} \Xi^{(E_g)}_{s,(1)}\nonumber\\
H_{4,(1)}^{(E_g)} &= \Phi^{(E_{g})}_{r,(1)}\Xi^{(E_g)}_{s,(1)}-\Phi^{(E_{g})}_{r,(2)}\Xi^{(E_g)}_{s,(2)},
&\qquad
H_{4,(2)}^{(E_g)} &=- \Phi^{(E_{g})}_{r,(1)}\Xi^{(E_g)}_{s,(2)}-\Phi^{(E_{g})}_{r,(2)}\Xi^{(E_g)}_{s,(1)}
\end{alignat}

and even more generally (all \irreps~ included):

\begin{alignat}{2}
\label{eq: Clebsch_Gordan2}
\Omega_1^{(A_{1g})}
&=
\Phi^{(A_{1g})}_{r} \Psi_s^{(A_{1g})}
&\nonumber\\
\Omega_2^{(A_{1g})}
&=
\Phi^{(A_{2g})}_{r} \Psi_s^{(A_{2g})}
&\nonumber\\
\Omega_3^{(A_{1g})}
&=
\Phi^{(E_{g})}_{r,(1)} \Psi_{s,(1)}^{(E_{g})}
+
\Phi^{(E_{g})}_{r,(2)} \Psi_{s, 2}^{(E_{g})}
&\nonumber\\[1ex]
\Omega_1^{(A_{2g})}
&=
\Phi^{(A_{2g})}_{r} \Psi_s^{(A_{1g})}
&\nonumber\\
\Omega_2^{(A_{2g})}
&=
\Phi^{(A_{1g})}_{r} \Psi_s^{(A_{2g})}
&\nonumber\\
\Omega_3^{(A_{2g})}
&=
\Phi^{(E_{g})}_{r,(1)} \Psi_{s,(2)}^{(E_{g})}
-
\Phi^{(E_{g})}_{r,(2)} \Psi_{s,(1)}^{(E_{g})}
&\nonumber\\[1ex]
\Omega_{1,(1)}^{(E_g)} &= \Phi^{(E_{g})}_{r,(1)}\Psi_s^{(A_{1g})},
&\qquad
\Omega_{1,(2)}^{(E_g)} &= \Phi^{(E_{g})}_{r,(2)}\Psi_s^{(A_{1g})},
\nonumber\\
\Omega_{2,(1)}^{(E_g)} &= \Phi^{(E_{g})}_{r,(2)}\Psi_s^{(A_{2g})},
&\qquad
\Omega_{2,(2)}^{(E_g)} &= -\Phi^{(E_{g})}_{r,(1)}\Psi_s^{(A_{2g})},
\nonumber\\
\Omega_{3,(1)}^{(E_g)} &= \Phi^{(A_{1g})}_{r} \Psi^{(E_g)}_{s,(1)},
&\qquad
\Omega_{3,(2)}^{(E_g)} &= \Phi^{(A_{1g})}_{r} \Psi^{(E_g)}_{s,(2)}\nonumber\\
\Omega_{4,(1)}^{(E_g)} &= \Phi^{(A_{2g})}_{r} \Psi^{(E_g)}_{s,(2)},
&\qquad
\Omega_{4,(2)}^{(E_g)} &= -\Phi^{(A_{2g})}_{r} \Psi^{(E_g)}_{s,(1)}\nonumber\\
\Omega_{5,(1)}^{(E_g)} &= \Phi^{(E_{g})}_{r,(1)}\Psi^{(E_g)}_{s,(1)}-\Phi^{(E_{g})}_{r,(2)}\Psi^{(E_g)}_{s,(2)},
&\qquad
\Omega_{5,2}^{(E_g)} &=- \Phi^{(E_{g})}_{r,(1)}\Psi^{(E_g)}_{s,(2)}-\Phi^{(E_{g})}_{r,(2)}\Psi^{(E_g)}_{s,(1)}
\end{alignat}

where, once again, I have omitted the repetition of $\Psi^{(\Gamma)}_s$ and $\Phi^{(\Gamma')}_r$ with the same \irrep~label.  When these are taken into account, there is a total of 60 mixed canonical polynomials in $\Xi^{(\Gamma)}_s$ and $\Phi^{(\Gamma')}_r$, consistent with the decomposition $\Gamma^{(\Phi)}\otimes \Gamma^{(\Xi)}=8 A_{1g}+ 12 A_{2g}+20 E_g$.  There is also a total of 150 mixed canonical polynomials in $\Psi^{(\Gamma)}_s$ and $\Phi^{(\Gamma')}_r$, consistent with the decomposition $\Gamma^{(\Phi)}\otimes \Gamma^{(\Psi)}=22 A_{1g}+ 28 A_{2g}+50 E_g$. \newline

The expression in Equation \ref{eq: Clebsch_Gordan2} can be re-written in general form as:

\begin{equation}
\Omega^{(\Gamma)}_{\alpha}= C^{(\Gamma, \Gamma', \Gamma'')}_{\alpha, \beta,\gamma} \Phi^{(\Gamma')}_{\beta} \Psi^{(\Gamma'')}_{\gamma} 
\end{equation}

  where the $C^{(\Gamma, \Gamma', \Gamma'')}_{\alpha, \beta,\gamma} $ are the usual Clebsch-Gordan coefficients for the relevant point group. Importantly, these coefficients are completely independent of the particular copy of the \irreps~ involved (subscripts $s$ and $r$). \newline
  
I will also employ mixed-variable canonical polynomials in $\vec{L}$ and the electric field $\vec{E}$ to write the fully canonical form of the magneto-electric tensor.
  
 \begin{alignat}{2}
\label{eq: E_L}
X_1^{(A_{1u})}
&=
L_z E_z
&\nonumber\\
X_2^{(A_{1u})}
&=
L_x E_x
+
L_y E_y
&\nonumber\\[1ex]
X^{(A_{2u})}
&=
L_x E_y
-
L_y E_x
&\nonumber\\[1ex]
X_{1,(1)}^{(E_u)} &= L_yE_z,
&\qquad
X_{1,(2)}^{(E_u)} &= -L_xE_z,
\nonumber\\
X_{2,(1)}^{(E_u)} &= L_z E_y,
&\qquad
X_{2,(2)}^{(E_u)} &= -L_z E_x\nonumber\\
X_{3,(1)}^{(E_u)} &= L_xE_x-L_yE_y,
&\qquad
X_{3,(2)}^{(E_u)} &=- L_xE_y-L_yE_x
\end{alignat}

\subsection{Canonical forms for bipartition \irreps~$A_{2g}$ and $A_{2u}$ of point group $\bar{3}m$}
\label{Sec: Canonical_vectors_3m}

 In the examples in the next sections, we will study in detail the cases $\chi_{bip}=A_{2g}$ (for $\alpha$-Fe$_2$O$_3$) and $\chi_{bip}=A_{2u}$ (for Cr$_2$O$_3$), which are the only two non-trivial 1D \irreps~ for point group $\hmn{-3m}$. As explained in Section \ref{sec: twisted projection}, canonical forms can be defined once and for all based on the crystal class and the bipartition representation, and can be later employed to construct any tensor on the canonical bases.  In particular, canonical vectors are axial vectors, since they represent spin indices.  Therefore, in order for a canonical vector to exist for the $\Gamma$ \irrep~block, $\Gamma \otimes \chi_{bip}$ must be contained in the axial-vector representations of the parent group, which for $\hmn{-3m}$ is $A_{2g} \oplus E_g$.  It follows immediately that the only allowed components are $\vec{V}^{(A_{1g/u})}$ and $\vec{V}^{(E_{g/u})}$, since $A_{1g} \otimes A_{2g}=A_{2g}$, $A_{1u} \otimes A_{2u}=A_{2g}$, $E_{g} \otimes A_{2g}=E_{g}$ and $E_{u} \otimes A_{2u}=E_{g}$ and are labelled according to the \irrep~of the symmetry-adapted polynomials. \newline

Likewise, one can construct other canonical matrix forms, which will later be useful to express tensors on partial canonical bases (for example, canonical in the order parameter and Cartesian in the wavevectors).  The canonical forms for $\chi_{bip}=A_{2g}$ and $\chi_{bip}=A_{2u}$ are the same (this is a general rule --- see Section \ref{Sec: Construction_Intertwiners}).  The procedure is a completely straightforward application of the projector operators, and was implemented in Mathematica using the `twisted' projector method described in Section \ref{sec: twisted projection}.  One obtains:  

\begin{description}

\item[Canonical vectors] $V^{(A_{1g/u})}$ and  the pair $V^{(E_{g/u})}_{(1)}, V^{(E_{g/u})}_{(2)}$ have 1 independent parameter each.

\begin{eqnarray}
\label{eq: wmag_canonicals}
V^{(A_{1g/u})}&=&\left(
\begin{smallmatrix}
 0  \\
 0 \\
a \\
\end{smallmatrix}
\right)\nonumber\\
V^{(A_{2g/u})}&=&\left(
\begin{smallmatrix}
 0  \\
 0 \\
0 \\
\end{smallmatrix}
\right)\nonumber\\
V^{(E_{g/u})}_{(1)}&=&\left(
\begin{smallmatrix}
 0  \\
 b \\
0 \\
\end{smallmatrix}
\right) \qquad V^{(E_{g/u})}_{(2)}=\left(
\begin{smallmatrix}
 -b  \\
 0 \\
0 \\
\end{smallmatrix}
\right)
\end{eqnarray}

\item[Canonical $3 \times 3$ matrices] these are useful either to express Cartesian $3 \times 3$ tensors on a symmetry-adapted basis of $\vec{L}$ polynomials (e.g., for the magneto-electric tensor).  $N^{(A_{1g/u})}$ has 1 independent parameter, $N^{(A_{2g/u})}$ has 2 independent parameters, $N^{(E_{g/u})}$ has 3 independent parameters.

\begin{equation}
\label{eq: alter_canonical_3}
\begin{aligned}
N^{(A_{1g/u})}&=\begin{pmatrix}
0 & a & 0\\[1ex]
-a & 0 & 0\\[1ex]
0 & 0 & 0
\end{pmatrix}\\
N^{(A_{2g/u})}&=
\begin{pmatrix}
b &0 & 0\\[1ex]
0& b & 0\\[1ex]
0 & 0 & c
\end{pmatrix}
\\
N^{(E_{g/u})}_{(1)}&=\begin{pmatrix}
0 & d & e\\[1ex]
d & 0 & 0\\[1ex]
f & 0 & 0
\end{pmatrix}, \qquad
N^{(E_{g/u})}_{(2)}=\begin{pmatrix}
d &0 & 0\\[1ex]
0& -d & e\\[1ex]
0 & f & 0
\end{pmatrix}
\end{aligned}
\end{equation}

\item[Canonical $3 \times 6$ matrices] these are useful to express Cartesian tensors in the Voigt form\footnote{In this work, the conventional engineering-strain factor of 2 is omitted from the Voigt representation of the shear components, consistently with the convention adopted in MTENSOR. This differs from the convention used in Ref.~\onlinecite{Radaelli2024}.}  --- for example, the piezo-magnetic tensor --- on a symmetry-adapted basis of $\vec{L}$ polynomials.  $M^{(A_{1g/u})}$ has 4 independent parameters, $M^{(A_{2g/u})}$ has 2 independent parameters, $M^{(E_{g/u})}$ have 6 independent parameters:

\begin{equation}
\label{Eq: Voigt_canonical}
\begin{aligned}
M^{(A_{1g/u})}
&=
\begin{pmatrix}
0 & 0 & 0 & 0 & a & b \\
b & -b & 0 & a & 0 & 0 \\
c & c & d & 0 & 0 & 0
\end{pmatrix},
\\[2ex]
M^{(A_{2g/u})}
&=
\begin{pmatrix}
e & -e & 0 & f & 0 & 0 \\
0 & 0 & 0 & 0 & -f & -e \\
0 & 0 & 0 & 0 & 0 & 0
\end{pmatrix},
\\[2ex]
M^{(E_{g/u})}_{(1)}
&=
\begin{pmatrix}
0 & 0 & 0 & 0 & g & -\dfrac{i-j}{2} \\
i & j & k & -g & 0 & 0 \\
l & -l & 0 & m & 0 & 0
\end{pmatrix}
\qquad
M^{(E_{g/u})}_{(2)}
=
\begin{pmatrix}
-j & -i & -k & -g & 0 & 0 \\
0 & 0 & 0 & 0 & -g & \dfrac{i-j}{2} \\
0 & 0 & 0 & 0 & -m & -l
\end{pmatrix}.
\end{aligned}
\end{equation}

\item[Canonical $3 \times 36$ matrices] these are useful to express Cartesian tensors in the double Voigt form  --- for example, the quadratic piezo-magnetic tensor --- on a symmetry-adapted basis of $\vec{L}$ polynomials. The canonical form  $R^{(A_{1g/u})}$ has 9 independent parameters, $R^{(A_{2g/u})}$ has 6 independent parameters, while $R^{(E_{g/u})}_{(1)}$ and $R^{(E_{g/u})}_{(2)}$ have 15 independent common parameters.  These matrices are rather large, and are reported in Appendix \ref{app: 3times36 matrices}.
\end{description}

\section{Examples: $\alpha$-\textsc{F\lowercase{e}$_2$O$_3$} and \textsc{C\lowercase{r}$_2$O$_3$}}

The pair of antiferromagnets $\alpha$-Fe$_2$O$_3$ and Cr$_2$O$_3$ was discussed by Dzyaloshinskii \cite{DZYALOSHINSKI1959, Dzyaloshinsky1958} who famously concluded that weak ferromagnetism and the magneto-electric effects are mutually exclusive for centrosymmetric structures.  Both compounds crystallise in the rhombohedral corundum structure, and the two ground-state magnetic structures, which were determined in the 1950s\cite{Shull1951,Brockhouse1953}  well before the development of the Bertaut--Izyumov formalism, are closely related, with the magnetic moments pointing along the crystallographic $c$-axis.  However, the alternation of magnetic moments on the four sites of the unit cell is different, and so is the action of the 12 operators of the group on the two magnetic sublattices.  In both cases, two-fold operations connect `up' with 'down' sites, and three-fold operations preserve the magnetic sites, which lie on the three-fold axis.  However, in  $\alpha$-Fe$_2$O$_3$, inversion connects `up'  with `up' sites while mirrors connect `up'  with `down' sites, while the opposite is true for Cr$_2$O$_3$.  On this basis, we can easily construct the corresponding bipartition  \irreps~ (the generators of the exchange multiplets) by assigning character $+1$ to operations that do not exchange sublattices and $-1$ for those that do.  As shown in  Table \ref{Table: Fe2O3_Cr2O3}, the two 1D representations are $A_{2g}$ for $\alpha$-Fe$_2$O$_3$ and $A_{2u}$ for Cr$_2$O$_3$.

Both  $\alpha$-Fe$_2$O$_3$ and Cr$_2$O$_3$ have accessible exchange multiplets although the anisotropy is much lower for $\alpha$-Fe$_2$O$_3$, so that all directions of $\vec{L}$ can be reached by changing temperature and/or applying a small magnetic field.  Cr$_2$O$_3$ does not possess zero-field weak magnetisation, and much larger magnetic fields are required to reach the spin flop transitions. 

The key symmetry information for these two compounds is summarised in Table \ref{Table: Fe2O3_Cr2O3}.  Each exchange multiplet comprises two distinct \irreps~--- $A_{1g}+E_g$ for $\alpha$-Fe$_2$O$_3$ and $A_{1u}+E_u$ for Cr$_2$O$_3$.  The $A_{1g}$ and $A_{1u}$ \irreps~correspond to the N\'eel vector along the $c$ axis and define a single MPG each.  The situation is slightly more complex for the $E_g/E_u$ \irreps, corresponding to the N\'eel vector in the plane orthogonal to $c$.  First, all operations corresponding to image matrices that are not real and diagonal are removed, yielding a 1D  `MPG irrep' $\chi_{MPG}$ for each direction of $\vec{L}$ within the $E_u$ doublet.  Then  the MPG symbol (Table \ref{Table: Fe2O3_Cr2O3}) is obtained by 'priming' the remaining operators that correspond to a $-1$ in $\chi_{MPG}$.  This is equivalent to an overall reversal of the N\'eel vector under the combined action of the permutation and the space symmetry.   For a completely general direction of $\vec{L}$, all operators in the MPG are lost except $\bar{1}$ or $\bar{1}'$.

\begin{table}[htbp]

\caption{\label{Table: Fe2O3_Cr2O3} Comparison of magnetic symmetry properties of $\alpha$-Fe$_2$O$_3$ and Cr$_2$O$_3$.  The 2-fold axis is parallel to the $(100)$ direction and equivalent.}
\centering
\renewcommand{\arraystretch}{1.2}
\begin{tabular}{|l|c|c|c|c|c|c|c|c|c|c|c|c|}
\hline
& \multicolumn{6}{c|}{$\alpha$-Fe$_2$O$_3$} & \multicolumn{6}{c|}{Cr$_2$O$_3$} \\
\hline\hline
Space group 
& \multicolumn{6}{c|}{$R\bar{3}c$} 
& \multicolumn{6}{c|}{$R\bar{3}c$} \\
\hline
bipartition irrep 
& \multicolumn{6}{c|}{$A_{2g}$} 
& \multicolumn{6}{c|}{$A_{2u}$} \\
\hline
Axial vector rep 
& \multicolumn{6}{c|}{$A_{2g}+E_g$} 
& \multicolumn{6}{c|}{$A_{2g}+E_g$} \\
\hline
Exchange multiplet 
& \multicolumn{6}{c|}{$A_{1g}+E_g$} 
& \multicolumn{6}{c|}{$A_{1u}+E_u$} \\
\hline
\hline
Operator classes 
& 1 & 3 & 2 & -1 & -3 & m 
& 1 & 3 & 2 & -1 & -3 & m \\
\hline
$\chi_{bip}$($A_{2g}$ or $A_{2u}$) 
& 1 & 1 & -1 & 1 & 1 & -1
& 1 & 1 & -1 & -1 & -1 & 1 \\
\hline
Spin group 
& \multicolumn{6}{c|}{$R\bar{3}^1\,2^2/c^2$}
& \multicolumn{6}{c|}{$R\bar{3}^2\,2^2/c^1$} \\
\hline
\multicolumn{13}{|c|}{$L \parallel 001$} \\
\hline
magnetic \textit{irrep} $\chi$
& 1 & 1 & 1 & 1 & 1 & 1
& 1 & 1 & 1 & -1 & -1 & -1 \\
\hline
MPG
& \multicolumn{6}{c|}{$\bar{3}\,2/m$}
& \multicolumn{6}{c|}{$\bar{3}'\,2/m'$} \\
\hline
\multicolumn{13}{|c|}{$L \parallel (100)$} \\
\hline
magnetic \textit{irrep} $\chi$
& 1 & $\times$  & -1 & 1 & $\times$  & -1
& 1 & $\times$  & -1 & -1 & $\times$  & 1 \\
\hline
MPG
& \multicolumn{6}{c|}{$2'/m'$}
& \multicolumn{6}{c|}{$2'/m$} \\
\hline
\multicolumn{13}{|c|}{$L \perp (100)$} \\
\hline
magnetic \textit{irrep} $\chi$
& 1 & $\times$ & 1 & 1 & $\times$  & 1
& 1 & $\times$  & 1 & -1 & $\times$  & -1 \\
\hline
MPG
& \multicolumn{6}{c|}{$2/m$}
& \multicolumn{6}{c|}{$2/m'$} \\
\hline
\multicolumn{13}{|c|}{$L$ generic} \\
\hline
magnetic \irrep~ $\chi$
& 1 & $\times$  & $\times$  & 1 & $\times$  & $\times$ 
& 1 & $\times$  & $\times$  & -1 & $\times$  & $\times$  \\
\hline
MPG
& \multicolumn{6}{c|}{$\bar{1}$}
& \multicolumn{6}{c|}{$\bar{1}'$} \\
\hline
\end{tabular}
\end{table}

\subsection{Tensorial properties }

A list of physical properties allowed in the exchange multiplets of $\alpha$-Fe$_2$O$_3$ and Cr$_2$O$_3$ is reported in Table \ref{Tab: corundum_tensors}, together with the Jahn symbols of the corresponding tensors up to rank 5.  These properties are always allowed for a generic orientation of the N\'eel vector, though they vanish for particular orientations (i.e., specific MPGs).  Since all MPGs of $\alpha$-Fe$_2$O$_3$ contain the inversion operator $\bar{1}$, all non-vanishing tensor forms must be parity-even.  Hence, allowed time-reversal-odd operators include the Jahn forms $aeV$, $aV^2$, $aeV[V^2]$ etc. Likewise, since all MPGs of Cr$_2$O$_3$ contain the inversion operator combined with time reversal ($\bar{1}')$ all time-reversal-odd tensor forms must be parity-odd, i.e., $aV$, $aeV^2$, $aV[V^2]$ etc.  Hence, the set of allowed time-reversal-odd operators is mutually exclusive to that of $\alpha$-Fe$_2$O$_3$.

\begin{table}[t]
\centering
\caption{\label{Tab: corundum_tensors}
Time-reversal-odd tensor properties generally allowed in the exchange multiplets of $\alpha$-Fe$_2$O$_3$
(spin point group $\bar{3}^1 m^2$) and Cr$_2$O$_3$
(spin point group $\bar{3}^2 m^1$). Some of these properties vanish for particular orientations of the N\'eel vector
(e.g., the magnetisation in $\alpha$-Fe$_2$O$_3$ for $\vec{L} \parallel c$)
}
\begin{tabular}{ccccc}
\toprule
\multicolumn{3}{c}{$\alpha$-Fe$_2$O$_3$} &
\multicolumn{2}{c}{Cr$_2$O$_3$} \\
\cmidrule(r){1-3}\cmidrule(l){4-5}
Rank & Jahn symbol & Physical properties &
Jahn symbol & Physical properties \\
\midrule
1 &
$aeV$ &
\tcell{Magnetisation; magnetocaloric effect; magnetothermal effect;
pyromagnetic effect} &
$aV$ &
\tcell{Polar toroidal moment; pyrotoroidic effect;
toroidal caloric effect} \\[0.5ex]

2 &
$aV^2$ &
\tcell{Electrotoroidic effect} &
$aeV^2$ &
\tcell{Magnetoelectric effect} \\[0.5ex]

3 &
$aeV[V^2]$ &
\tcell{Piezomagnetic tensor; second-order magnetoelectric effect;
magnetic birefringence} &
$a[V^2]V$ &
\tcell{Piezotoroidic effect; spontaneous gyrotropic birefringence} \\[0.5ex]

4 &
-- &
\tcell{--} &
$aeV^2[V^2]$ &
\tcell{Flexomagnetic effect} \\[0.5ex]

5 &
$aeV[V^4]$ &
\tcell{Quartic piezomagnetism; quartic altermagnetism} &
-- &
\tcell{--} \\
\bottomrule
\end{tabular}
\end{table}

\subsection{Construction of the intertwiners}
\label{Sec: Construction_Intertwiners}

Starting from a physical tensor, say $aeV[V^2]$ in $\alpha$-Fe$_2$O$_3$, to construct the intertwiner one must remove the time-reversal symbol $a$ (which is contained in the odd power of $\vec{L}$) and add the symbol corresponding to the power of $\vec{L}$ that is being intertwined ($eV$ for the linear term, $e[V^3]$ for the cubic term etc.).  When allowed, the absolute (co-rotating) intertwiner $\bm{\mathcal{A}}$ is obtained from the  first-order intertwiner $\bm{\mathcal{I}^{(1)}}$ by removing the spin indices (to be assigned to the $\delta$ function or to other spherically symmetric tensor). For example, from the tensor form for the piezomagnetic effect $aeV[V^2]$, one obtains

\begin{eqnarray}
\label{eq: intertwiner_derivation}
\bm{\mathcal{I}}^{(1)}&=&\cancel{a} eV[V^2] \, e V = V^2[V^2]\nonumber\\
\bm{\mathcal{I}}^{(3)}&=&\cancel{a} eV[V^2] \, e [V^3] = V[V^2][V^3]\nonumber\\
\bm{\mathcal{A}}&=&\cancel{V^2}[V^2]
\end{eqnarray}

Once again, I emphasise that the intertwiners must be symmetrised with the bipartition \irrep~(not with the totally symmetric representation of some MPG).  One interesting observation is that  corresponding pairs of tensors for $\alpha$-Fe$_2$O$_3$ and Cr$_2$O$_3$  in the same row of Table \ref{Tab: corundum_tensors} (for example, $aeV[V^2]$ and $aV[V^2]$) have \emph{identical intertwiners}.  In fact, both the tensors and the symmetrising $\chi_{bip}$ have opposite parities, yielding identical intertwiners in the two cases.  For example, the magnetoelectric tensor in Cr$_2$O$_3$ has the same intertwiners as the electro-toroidic tensor in Fe$_2$O$_3$, and the same is true of the magnetisation (Fe$_2$O$_3$) and polar toroidal moment (Cr$_2$O$_3$).

\subsection{Cartesian and canonical intertwiners: general approach}

In the following sections, I will present intertwiners relating to several physical properties of $\alpha$-Fe$_2$O$_3$ and Cr$_2$O$_3$ in both Cartesian and canonical forms.  The following general approach will be followed:

\begin{enumerate}
\item For simple tensors, I will first present the Cartesian form of the intertwiners, which can be obtained very simply by adapting the functionalities of the Bilbao Crystallographic Server (program MTENSOR\cite{PerezMato2015, Gallego2019}) as explained in Appendix \ref{App: MTENSOR_Adaptation}.  Since this construction is straightforward and easily reproducible, I will omit it for more complex tensors.
\item I will then construct several intertwiner forms on partial canonical bases (e.g., Cartesian in $\vec{k}$ and canonical in $\vec{L}$), which seem to be convenient for different purposes (which I shall explain).  The connection between the coefficients of the fully Cartesian tensors and those of the canonical forms will be shown explicitly for simple tensors.  It is calculated for more complex tensors by a Mathematica code, which is available as part of the Supplementary Information \cite{MyPaperSI}.
\item In some cases, I will also present the tensors expressed in terms of the fully canonical vector intertwiners.  A possible computational approach to exploit the fully canonical intertwiner approach will be discussed separately in Appendix \ref{App: Fully_Canonical_CG}.
\end{enumerate}

\subsection{Weak magnetisation and the $\vec{T}$ vectors in $\alpha$-Fe$_2$O$_3$}

The weak magnetisation (Jahn symbol $aeV$) is  a fundamental property of $\alpha$-Fe$_2$O$_3$ and other antiferromagnets, and is key to their functional properties because it enables one to control  domains.  It has been known since the 1950s that the absence of the $\bar{1}'$ operator, i.e., the inversion combined with time reversal, is a prerequisite for weak ferromagnetism.  In collinear zone-centre antiferromagnets, the absence of $\bar{1}'$  also characterises altermagnets and distinguishes them from other antiferromagnets.  In addition to the weak magnetisation $\vec{m}$ (a TRO axial vector), altermagnets admit in general other axial vectors that  are not necessarily collinear with $\vec{m}$.  These include the XMCD and MOKE vectors (generally denoted $\vec{T}$), which define the antisymmetric part of the optical conductivity tensor $\bm{\sigma}$ through the formula:

\begin{equation}
\sigma_{ij}=\epsilon_{ijk} T_k
\end{equation} 

Given what was discussed in the previous paragraph, it is clear that such axial vectors cannot have a co-rotating component.  This is immediately obvious if one derives the Jahn symbols for the intertwiners, as in Equation \ref{eq: intertwiner_derivation}: 

\begin{eqnarray}
\label{eq: intertwiner_derivation_wmag}
\bm{\mathcal{I}}^{(1)}&=&\cancel{a} eV\, e V = V^2\nonumber\\
\bm{\mathcal{I}}^{(3)}&=&\cancel{a} eV\, e [V^3] = V[V^3]\nonumber\\
\bm{\mathcal{A}}&=&\cancel{V^2} = [0]
\end{eqnarray}

Here, I derive the expression of a generic TRO axial vector in $\alpha$-Fe$_2$O$_3$ for a general direction of the order parameter $\vec{L}$ and up to third order.

\subsubsection{Cartesian forms}
To first and third order in $\vec{L}$, the symmetrised intertwiner tensors (from MTENSOR) have the following form:

\begin{equation}
\label{Eq: linear_weak}
\bm{\mathcal{I}}^{(1)}=\left(\begin{smallmatrix}
 0 & c_{12}& 0 \\
-c_{12}& 0 & 0 \\
 0 & 0 & 0 \\
\end{smallmatrix}\right)
\end{equation}

(one independent parameter), while at the third order the Cartesian intertwiner reads:
 
\begin{equation}
\label{Eq: cubic_weak}
\bm{\mathcal{I}}^{(3)}=
\renewcommand{\arraystretch}{1.15}
\begin{array}{c|cccccc|cccccc|cccccc}
 &11&21&31&41&51&61&
 12&22&32&42&52&62&
 13&23&33&43&53&63\\
\hline
1&
0&0&0&0&
c_{113}&c_{112}&
c_{112}&3c_{112}&c_{132}&-c_{113}&0&0&
c_{113}&-c_{113}&0&c_{132}&0&0
\\
2&
-3c_{112}&-c_{112}&-c_{132}&-c_{113}&
0&0&
0&0&0&0&
-c_{113}&-c_{112}&
0&0&0&0&
-c_{132}&-c_{113}
\\
3&
c_{311}&-c_{311}&0&0&
0&0&
0&0&0&0&
0&-c_{311}&
0&0&0&0&
0&0
\end{array}
\end{equation}

where this last tensor has four independent parameters and employs the 6-index Voigt notation with the Voigt array defined as

\begin{equation}
\mathbf{L}_{\mathrm V}^{(2)}
=
(L_x^2,\,
L_y^2,\,
L_z^2,\,
2L_yL_z,\,
2L_xL_z,\,
2L_xL_y)^{\mathrm T}.
\end{equation}

\subsubsection{Canonical forms}

As explained in Section \ref{Sec: basis transformation}, one can perform a transformation from the Cartesian Voigt form to the canonical form, expressed at the linear order in the $\left(L_x, L_y, L_z\right)$ basis (which is already canonical) or, at the third order, on the $\Phi$ basis (Equation \ref{Eq: l_cube_poly}): \newline 

Defining the \irrep~canonical vectors as in Equation \ref{eq: wmag_canonicals}, for the linear tensor in $\vec{L}$ one obtains:

\begin{equation}
\vec{T}^{(1)}   =V^{E_{g/u}}_{(1)}\, L_x +
V^{E_{g/u}}_{(2)}\, L_y
\end{equation}

where the connection with the Cartesian tensor coefficient in Equation \ref{Eq: linear_weak} is:

\begin{equation}
V^{(E_{g/u})}_{(1)}=\left(
\begin{smallmatrix}
 0  \\
 c_{12} \\
0 \\
\end{smallmatrix}
\right)  \qquad V^{(E_{g/u})}_{(2)}=\left(
\begin{smallmatrix}
 -c_{12}  \\
 0 \\
0 \\
\end{smallmatrix}
\right)
\end{equation}
\newline

Likewise, for the third order in $\vec{L}$ and expanding on the symmetry-adapted basis one obtains the following form:

\begin{equation}
\label{eq: third_order_wmag}
\vec{T}^{(3)}=
V^{(A_{1g/u})}\Phi^{(A_{1g})}
+\sum_{r=1}^{3}
\sum_{\alpha=1}^{2}
V^{E_{g/u}}_{r,(\alpha)}\,
\Phi^{E_{g/u}}_{r,(\alpha)}.
\end{equation}

where the terms in $A_{2g}$ have been omitted since $V^{(A_{2g/u})}=\vec{0}$, and the constants to be inserted in the other canonical vectors (Equation \ref{eq: wmag_canonicals}) are related to the elements of the Cartesian tensor in Equation \ref{Eq: cubic_weak} as:

\begin{eqnarray}
a&=&c_{311}\nonumber\\
b_1&=&-\frac{3}{2} (c_{112} +c_{132})\nonumber\\
b_2&=&-\frac{3}{2} (c_{112} -c_{132})\nonumber\\
 b_3&=& c_{113}
\end{eqnarray}

These results are fully consistent with those presented in Ref. \onlinecite{mallon2026revealingoriginxmcdaltermagnet}.  In particular, for $\vec{L} \perp c$ the formulas above retrieve the 120$^\circ$--periodic term in $T_z$ through the totally-symmetric polynomial $\Phi^{(A_{1g})}= L_x\!\left(L_x^2 - 3L_y^2\right)$.
\newline

I would like to draw attention to the extreme simplicity of this approach:  to deduce the full form of Equation \ref{eq: third_order_wmag}, all one needs to know is the form of the canonical vectors in Equation \ref{eq: wmag_canonicals}.  This is generally the case for any tensor form, as explained in the examples here below, but it is particularly simple here because the weak magnetisation is not contracted with other physical variables, and the expression in Equation \ref{eq: third_order_wmag} is fully canonical.

\subsection{Spin-texture tensors for $\alpha$-Fe$_2$O$_3$}

Spin textures defined by quadratic tensors in $\vec{k}$ are extensively discussed in the altermagnetism literature (see Ref. \onlinecite{Radaelli2024} and citations therein), and have the same form as the linear piezomagnetic tensor. At linear and quadratic orders in $\vec{L}$ the corresponding intertwiners have Jahn symbols $V^2[V^2]$ (8 independent parameters) and $V[V^2][V^3]$ (28 independent parameters).  This arises from the decomposition of the generic tensors $V^2[V^2]=10A_{1g}\oplus 8 A_{2g}\oplus 18 E_g$,  $V[V^2][V^3]=32A_{1g}\oplus 28 A_{2g}\oplus 60 E_g$ and the fact that the intertwiners are projected onto $\chi_{bip}=A_{2g}$.\newline

Spin textures defined by quartic tensors in $\vec{k}$ have also been discussed in the context of $g$-wave altermagnetism, and have the same form as the quadratic (in the strain) piezomagnetic tensor.  At linear and quadratic orders in $\vec{L}$ the intertwiners have Jahn symbols $V^2[V^4]=24A_{1g}\oplus 21 A_{2g}\oplus 45 E_g$ (21 independent parameters when projected on $\chi_{bip}$) and $V[V^4][V^3]=78A_{1g}\oplus 72 A_{2g}\oplus 150 E_g$ (72 independent parameters when projected on $\chi_{bip}$).  The 6\textsuperscript{th}-rank linear intertwiner  $V^2[V^4]$ can also be calculated using MTENSOR, but $V[V^4][V^3]$ (8\textsuperscript{th}-rank) is too large for MTENSOR.\newline

The fully Cartesian forms of these intertwiners up to rank 6 have been constructed using the MTENSOR programme, as explained in Appendix \ref{App: MTENSOR_Adaptation}.  I then projected the Cartesian tensors onto the appropriate canonical bases using the Mathematica code, and generated a correspondence map between the Bilbao and the canonical form parameters.  However, the fully Cartesian parametrisation for intertwiners is not especially useful or even physical, so it will be omitted.  The Mathematica code generates all the parameter maps, and is available as part of the Supplementary Information.\cite{MyPaperSI}

\subsubsection{Representation on the $\vec{L}$ canonical basis}

As already mentioned, this is possibly the most useful representation, because it enables one to construct tensors in the familiar Cartesian form to arbitrary powers of the order parameters, provided that the corresponding symmetry-adapted polynomials in $\vec{L}$ have been calculated.

\begin{description}

\item[Quadratic in $\vec{k}$: linear piezomagnetism and quadratic spin textures] as previously observed, the $(L_x, L_y, L_z)$ is already canonical since $(L_x, L_y)$ transform with $E_g$ and $L_z$ transforms with $A_{2g}$.  The cubic canonical basis functions are the $\Phi^{(\Gamma)}$ polynomials in Equation \ref{Eq: l_cube_poly}.  Consequently one may write:

\begin{equation}
\mathbf{s}(\mathbf{k},\mathbf{L})
=\left(\ten{T}^{(1)}(\vec{L}) +\ten{T}^{(3)}(\vec{L})+\dots\right)  \,\mathbf{K}_{\mathrm V}^{(2)}
\end{equation}

with 

\begin{equation}
\mathbf{K}_{\mathrm V}^{(2)}
=
(k_x^2,\,
k_y^2,\,
k_z^2,\,
2k_yk_z,\,
2k_xk_z,\,
2k_xk_y)^{\mathrm T}
\end{equation}

and

\begin{equation}
\ten{T}^{(1)}(\vec{L})=
M^{(A_{2g/u})} L_z
+
M^{(E_g/u)}_{(1)} L_x+M^{(E_g/u)}_{(2)}L_y
\end{equation}

\begin{equation}
\ten{T}^{(3)}(\vec{L}) =
M^{(A_{1g/u})}_1 \Phi^{(A_{1g})}
+
\sum_{r=1}^{3} M^{(A_{2g/u})}_{r}\Phi^{(A_{2g})}_{r}
+
\sum_{r=1}^{3}\sum_{\alpha=1}^{2}M^{(E_{g/u})}_{r,(\alpha)}\Phi^{(E_{g/u})}_{r,(\alpha)}
\end{equation}

and the $M$ matrices from Equation \ref{Eq: Voigt_canonical}.  At the linear order in $\vec{L}$,  the 8 parameters of the fully Cartesian tensor are combined into $M^{(A_{2g/u})}$ (2 independent parameters) and $M^{(E_{g/u})}$ (6 independent parameters).  At the cubic order, the 28 parameters of the fully Cartesian tensor are combined into $M^{(A_{1g/u})}$ (4 independent parameters), 3 copies of $M^{(A_{2g/u})}$ (2 independent parameters each) and 3 copies of $M^{(E_{g/u})}$  (6 independent parameters each).

\item[Quartic order in $\vec{k}$: quadratic piezomagnetism and quartic spin textures] these can be written as:

\begin{equation}
\mathbf{s}(\mathbf{k},\mathbf{L})
=\left(\ten{T}^{(1)}(\vec{L}) +\ten{T}^{(3)}(\vec{L})+\dots\right)  \,\mathbf{K}_{\mathrm V}^{(4)}
\end{equation}

with 

\begin{equation}
\mathbf{K}_{\mathrm V}^{(4)}
= \mathbf{K}_{\mathrm V}^{(2)} \otimes \mathbf{K}_{\mathrm V}^{(2)}
\end{equation}

and

\begin{equation}
\ten{T}^{(1)}(\vec{L})=
R^{(A_{2g/u})} L_z
+
R^{(E_g/u)}_{(1)} L_x+R^{(E_g/u)}_{(2)}L_y
\end{equation}

\begin{equation}
\ten{T}^{(3)}(\vec{L}) =
R^{(A_{1g})}\Phi^{(A_{1g})}
+
\sum_{r=1}^{3} R^{(A_{2g})}_{r}\Phi^{(A_{2g})}_{r}
+
\sum_{r=1}^{3}\sum_{\alpha=1}^{2}R^{(E_g)}_{r,(\alpha)}\Phi^{(E_g)}_{r,(\alpha)}
\end{equation}

and the $R$ matrices are discussed in Section \ref{Sec: Canonical_vectors_3m}.  Since $R^{(A_{1g/u})}$ has 9 independent parameters, $R^{(A_{2g/u})}$ has 6 independent parameters, while $R^{(E_{g/u})}_{(1)}$ and $R^{(E_{g/u})}_{(2)}$ have 15 independent common parameters, one finds that $\ten{T}^{(1)}(\vec{L})$ has 21 parameters and $\ten{T}^{(3)}(\vec{L)}$, has 72 parameters, exactly as in the fully Cartesian representations.  Building the complex $\ten{T}^{(3)}(\vec{L)}$ tensor is now completely straightforward:  one simply needs to write multiple copies of the canonical matrices (each with unique parameters) and assign each of the copies to the appropriate harmonic component in $\vec{L}$. Projections on particular directions or planes in the order parameter space  to match experimental geometries is also straightforward.

\end{description}

\subsubsection{Canonical representation on the $\vec{k}$ canonical bases: SOC-free altermagnetism}

At linear order in $\vec{L}$, it is convenient to re-write the same tensors on the $k_ik_j$ or $k_ik_jk_lk_m$  canonical bases, because this allows a clean separation between the co-rotating (SOC-free) and non-co-rotating parts of the tensor.

At quadratic and quartic order in $\vec{k}$, one employs the $\Xi$ and $\Psi$ polynomials in Equations \ref{eq: Xi_poly} and \ref{eq: Psi_poly}, respectively and writes the following expansion in powers of $\vec{k}$:  
\begin{equation}
\mathbf{s}(\mathbf{k},\mathbf{L})
=
\left(
\ten{S}^{(2)}(\vec{k})+\ten{S}^{(4)}(\vec{k})+\dots
\right)(L_x, L_y,L_z)^{\mathrm T}
\end{equation}

with 

\begin{equation}
\ten{S}^{(2)}(\vec{k})
=
\sum_{r=1}^{2} N^{(A_{1g/u})}_{r} \Xi^{(A_{1g})}_{r}+\sum_{r=1}^{2}\sum_{\alpha=1}^{2}N^{(E_{g/u)}}_{r,(\alpha)}\Xi^{(E_g)}_{r,(\alpha)}
\end{equation}

\begin{equation}
\ten{S}^{(4)}(\vec{k})
=
\sum_{r=1}^{4} N^{(A_{1g})}_{r} \Psi^{(A_{1g/u})}_{r}+N^{(A_{2g/u})}_{1} \Psi^{(A_{2g})}_{1}+\sum_{r=1}^{5}\sum_{\alpha=1}^{2}N^{(E_{g/u})}_{r,(\alpha)}\Psi^{(E_g)}_{r,(\alpha)}
\end{equation}

For $\ten{S}^{(2)}(\vec{k})$, the 8 parameters of the fully Cartesian tensors are combined into two copies of $N^{(A_{1g/u})}$ (1 independent parameter each) and two copies of $N^{(E_{g/u})}$ (3 independent parameters each).  For $\ten{S}^{(4)}(\vec{k})$, the 21 parameters of the fully Cartesian tensors are combined into 4 copies of $N^{(A_{1g/u})}$ (1 independent parameter each) one copy of  $N^{(A_{2g/u})}$ (2 parameters) and 5 copies of $N^{(E_{g/u})}$ (3 independent parameters each).\newline

From this we can immediately see that there is no co-rotating component at quadratic order in $\vec{k}$ and that there is a single co-rotating component at quartic order. In fact, the only $N$ matrix that is not traceless is $N^{(A_{2g})}$, which appears zero times in $\ten{S}^{(2)}(\vec{k})$ and one time in $\ten{S}^{(4)}(\vec{k})$.  As previously noted in \ref{sec: canonical_absolute}, only the symmetry-adapted polynomial components transforming according to $\chi_{bip}$ contribute to absolute intertwiners, and there is no term in $A_{2g}$ in the $\Xi$ polynomial decomposition.  \newline

The absence of a co-rotating component in $\ten{S}^{(2)}(\vec{k})$  can also be shown by observing that all the elements of the Cartesian absolute intertwiner $\bm{\mathcal{A}}$ tensor (Jahn symbol $[V^2]$) in the co-rotating expression:

\begin{equation}
s^{(1)}_i(\mathbf{k},\mathbf{L})
=\delta_{ij}\mathcal{A}_{kl} L_jk_kk_l
\end{equation}

are zero.\newline

For $\ten{S}^{(4)}(\vec{k})$ ,  the co-rotating textures are:

\begin{equation}
\label{Eq: spin_corotating_gwave}
s^{cor}_i(\mathbf{k},\mathbf{L})
= \tau \delta_{ij} L_j \Psi^{(A_{2g})}_1=\tau L_i \,k_z k_x\!\left(k_x^2-3k_y^2\right) 
\end{equation}

where $\tau=Tr\left(N^{(A_{2g})}\right)$.  This is the well-known $g$-wave altermagnetic texture, again, as expected.
\newline

One can also obtain the same  by symmetrising  directly the Cartesian absolute intertwiner $\bm{\mathcal{A}}$ tensor, which has the expression (from MTENSOR):

\begin{equation}
\begin{array}{c|cccccc}
 & 1 & 2 & 3 & 4 & 5 & 6 \\
\hline
1 & 0 & 0 & 0 & 0 & c_{15} & 0 \\
2 & 0 & 0 & 0 & 0 & -c_{15} & 0 \\
3 & 0 & 0 & 0 & 0 & 0 & 0 \\
4 & 0 & 0 & 0 & 0 & 0 & -c_{15} \\
5 & c_{15} & -c_{15} & 0 & 0 & 0 & 0 \\
6 & 0 & 0 & 0 & -c_{15} & 0 & 0
\end{array}
\end{equation}

where Voigt indices were used for both dimensions.  Double contraction with $\mathbf{K}_{\mathrm V}^{(2)}$ yields:

\begin{equation}
s^{cor}_i(\mathbf{k},\mathbf{L})
= 4c_{15}  \delta_{ij} L_j \, k_z k_x\!\left(k_x^2-3k_y^2\right) 
\end{equation}

which is identical to the previous expression with $\tau\equiv4c_{15}$.

\subsubsection{Canonical vectors on mixed canonical bases}
As discussed in Section \ref{Sec: Mixed Canonical Bases}, one can introduce  mixed symmetry-adapted polynomial bases for both $\vec{L}$ and $\vec{k}$.  These are the simplest possible decompositions, since all the complexity is in the construction of the mixed-variable polynomials (independent of magnetism).  With the mixed canonical polynomial in Equations \ref{eq: Clebsch_Gordan}, \ref{eq: Clebsch_Gordan1b} and \ref{eq: Clebsch_Gordan2}, one writes: 

\begin{equation}
\mathbf{s}^{(1, 2)}(\mathbf{k},\mathbf{L})
=
\sum_{r=1}^{2} V^{(A_{1g})}_{r} Z^{(A_{1g})}_{r}+\sum_{r=1}^{6}\sum_{\alpha=1}^{2}V^{(E_g)}_{r,(\alpha)}Z^{(E_g)}_{r,(\alpha)}
\end{equation}

\begin{equation}
\mathbf{s}^{(3, 2)}(\mathbf{k},\mathbf{L})
=
\sum_{r=1}^{8} V^{(A_{1g})}_{r} H^{(A_{1g})}_{r}+\sum_{r=1}^{20}\sum_{\alpha=1}^{2}V^{(E_g)}_{r,(\alpha)}H^{(E_g)}_{r,(\alpha)}
\end{equation}

\begin{equation}
\mathbf{s}^{(3, 4)}(\mathbf{k},\mathbf{L})
=
\sum_{r=1}^{22} V^{(A_{1g})}_{r} \Omega^{(A_{1g})}_{r}+\sum_{r=1}^{50}\sum_{\alpha=1}^{2}V^{(E_g)}_{r,(\alpha)}\Omega^{(E_g)}_{r,(\alpha)}
\end{equation}

where the notation $s^{(i,j)}$ indicates $i$\textsuperscript{th} order in $\vec{L}$ and $j$\textsuperscript{th} order in $\vec{k}$ , and the terms in $A_{2g}$ have been omitted because $V^{(A_{2g})}= \vec{0}$.  Since each of the copies of $V^{(\Gamma)}$ has one free parameter (the same parameter $b$ for the two $E_g$ vectors), one can once again  retrieve the correct number of parameters in the fully Cartesian tensors (8, 28 and 72, respectively) from the \irrep~decomposition presented in Section \ref{sec: mixed_var}.\newline

\subsection{The magnetoelectric tensor of \textsc{C\lowercase{r}$_2$O$_3$}}

The intertwiner framework is by no means limited to altermagnets, and can be employed exactly in the same way to systems that are $PT$ symmetric.  As an example, I provide a full expansion of the magnetoelectric tensor $\ten{\alpha}$ of Cr$_2$O$_3$ up to the third order in $\vec{L}$, and demonstrate that $\ten{\alpha}(\vec{L})$ coincides with the tensors symmetrised with the MPG in specific directions of the N\'eel vector.   The direct form of the tensor, which connects the induced electrical polarisation $\vec P$ to the applied magnetic field $\vec H$ through the formula $P_i=\alpha_{ij}H_j$, will be discussed first.  This will demonstrate that the general framework is still applicable in cases in which the spin indices are contracted rather than free.  Prior to this, I will provide a comparison of the Landau--Lifshitz expansion of the magneto-electric coefficients with experimental literature data, a comparison which validates the principles discussed throughout this paper.

\subsubsection{The Landau--Lifshitz expansion: comparison with experimental data}

\begin{figure}[!h]
\centering
\includegraphics[scale=0.5]{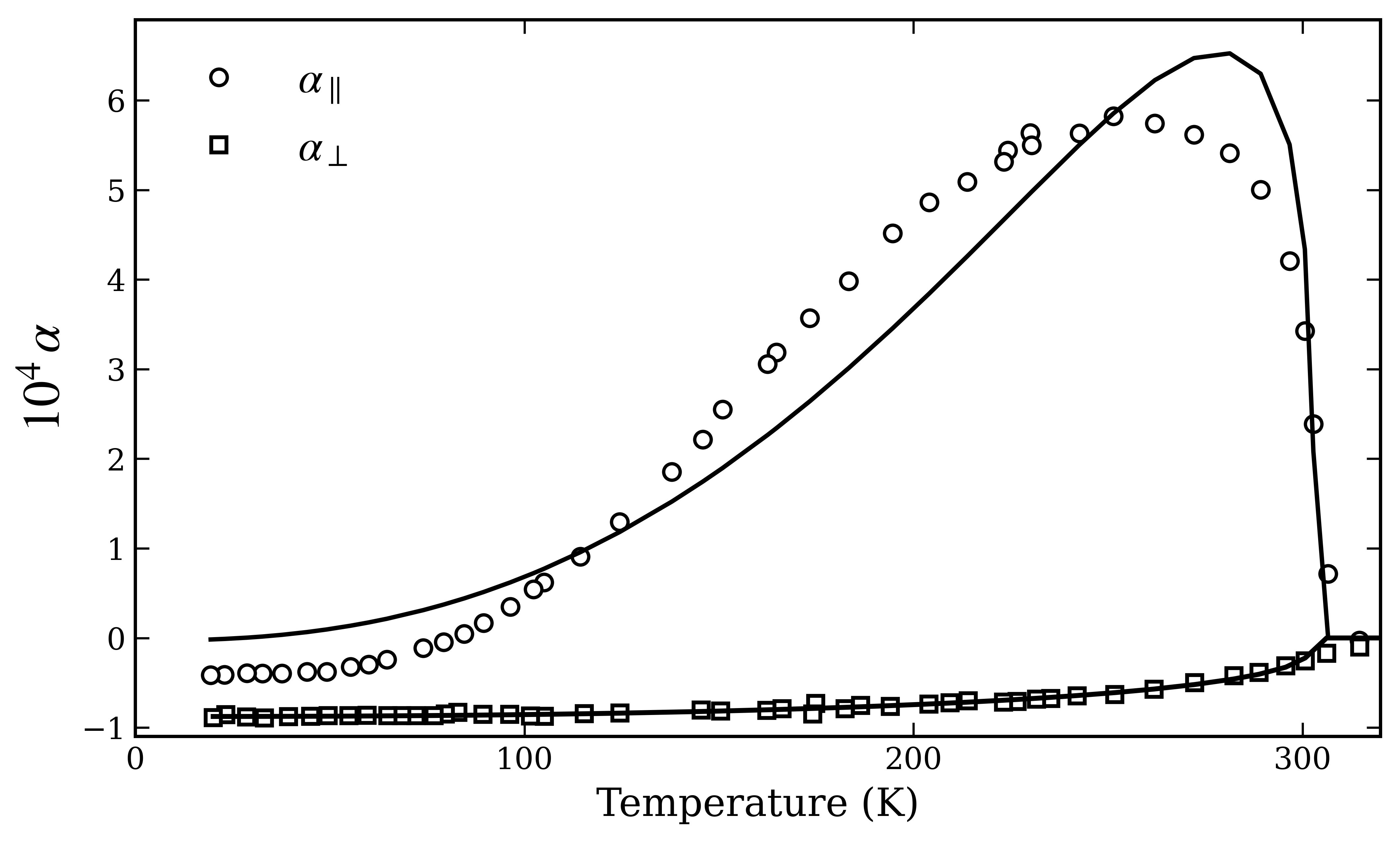}
\caption{\label{fig:ME_fit}
Temperature dependence of the parallel ($\alpha_{\parallel}$, circles) and
perpendicular ($\alpha_{\perp}$, squares) components of the linear
magnetoelectric tensor of Cr$_2$O$_3$. The experimental magnetoelectric data
are taken from Astrov \textit{et al.}~\cite{Astrov1961}. The solid curves are
simultaneous least-squares fits using the symmetry-allowed expansion
$\alpha=aL+bL^3$, where the antiferromagnetic order parameter is represented by
the phenomenological expression
$L(T)=L_0\left[1-\left(T/T_N\right)^{\alpha_L}\right]^{\beta}$,
fitted to the neutron-diffraction measurements of Shaked
\textit{et al.}~\cite{Shaked1968}. Values of $L(T)$ at the temperatures of the
magnetoelectric measurements were obtained by interpolation of this expression
before fitting the coefficients $a$ and $b$. The fitted coefficients are reported in the text.}
\end{figure}

Unlike the previous cases, the experimental temperature dependence of both the order parameter\cite{Shaked1968} and the  magneto-electric coefficients\cite{Astrov1961} has long been established experimentally for the equilibrium magnetic structure ($\vec{L} \parallel c$), allowing a direct comparison to be made with the Landau--Lifshitz expansion.  With this orientation of $\vec{L}$, the magnetoelectric tensor is diagonal, with two independent parameters $\alpha_\parallel$ and $\alpha_\perp$ corresponding to the magnetic field being oriented parallel/perpendicular (respectively) to the $c$ axis.
Figure~\ref{fig:ME_fit} compares the cubic Landau--Lifshitz expansion with the
temperature-dependent magnetoelectric measurements of Astrov
\textit{et al.}~\cite{Astrov1961}. Since the expansion is expressed in powers
of the antiferromagnetic order parameter rather than temperature, the order
parameter was first obtained from the neutron-diffraction measurements of
Shaked \textit{et al.}~\cite{Shaked1968}. The experimental order parameter was
represented by the phenomenological form

\begin{equation}
L(T)=L_0\left[1-\left(\frac{T}{T_N}\right)^{p}\right]^{\beta},
\label{eq:Lfit}
\end{equation}

where $L_0$, $T_N$, $p$, and $\beta$ were determined by least-squares fitting
to the neutron data. The resulting expression provides a smooth interpolation
from which the values of $L(T)$ were evaluated at the temperatures of the
magnetoelectric measurements.

The magnetoelectric coefficients were then fitted as functions of the
interpolated order parameter using the lowest-order symmetry-allowed expansion,

\begin{equation}
\alpha = aL+bL^3.
\label{eq:alphafit}
\end{equation}

A simultaneous least-squares fit was performed for the parallel and
perpendicular tensor components while constraining both datasets to share a
common N\'eel temperature. The fitted coefficients are

\begin{align}
\alpha_{\parallel}(L)
   &= a_{\parallel}L+b_{\parallel}L^3, \nonumber\\
a_{\parallel}
   &= 16.9(5) \qquad
b_{\parallel}
   =-16.8(4),\\[1ex]
\alpha_{\perp}(L)
   &= a_{\perp}L+b_{\perp}L^3, \nonumber\\
a_{\perp}
   &=-0.8(4),\qquad
b_{\perp}
   = -0.1(1)\nonumber\\
T_N&=302.3(4) K
\end{align}

where $L$ is the reduced order parameter ($0 \le L \le 1$) and the quoted parameter uncertainties are one-standard-deviation estimates obtained from the covariance matrix of the unweighted least-squares fit, using the residual variance to estimate the noise level.\newline As shown in Fig.~\ref{fig:ME_fit}, the cubic expansion qualitatively reproduces the overall
temperature dependence of both tensor components over the
entire ordered phase. The systematic deviations, which are visible at intermediate
temperatures, particularly for $\alpha_{\parallel}$, suggest that
higher-order odd powers of $L$ may also contribute.\newline

To assess the importance of such terms, the data were also fitted using

\begin{equation}
\alpha=aL+bL^3+cL^5.
\end{equation}

This extension yields an almost exact representation of the experimental data.
However, the fitted cubic and quintic coefficients are found to be strongly
correlated, indicating that the available measurements do not uniquely
determine the higher-order contribution. From this analysis one can therefore conclude that the third-order
expansion provides the minimal phenomenological description supported by the
present experimental data, while remaining fully consistent with the symmetry
analysis developed in this work.


\subsubsection{Construction of covariant magnetoelectric tensors}

As usual, we construct the third-order (cubic) tensor directly and extract the first order term as we have done previously

For the linear and cubic terms in $\vec{L}$ one obtains:

\begin{eqnarray}
\label{eq: ME_tensor_linear_compact}
\ten{\alpha}^{(1)}&=& N^{(A_{2g/u})}L_z
+
N^{(E_{g/u})}_{1}L_x+N^{(E_{g/u})}_{2}L_y\nonumber\\
&=&\left(\begin{smallmatrix}
 b L_z+d L_y&d L_x& e L_x \\
d L_x& b L_z-d L_y & e L_y \\
 f L_x& f L_y & c  L_z \\
\end{smallmatrix}\right)
\end{eqnarray}

Note that the linear tensor is not completely general in that it has only 5 parameters instead of 9.  The most general form is only retrieved at the 3\textsuperscript{rd} order:

\begin{equation}
\label{eq: ME_tensor_cubic_compact}
\ten{\alpha}^{(3)}=N^{(A_{1g/u})}\Phi^{(A_{1g})}
+
\sum_{r=1}^{3} N^{(A_{2g/u})}_{r}\Phi^{(A_{2g})}_{r}
+
\sum_{r=1}^{3}\sum_{\alpha=1}^{2}N^{(E_{g/u})}_{r,(\alpha)}\Phi^{(E_g)}_{r,(\alpha)}
\end{equation}

where the matrices in Equation \ref{eq: ME_tensor_linear_compact}  and \ref{eq: ME_tensor_cubic_compact} are the usual canonical matrices in Equation \ref{eq: alter_canonical_3} (which are the \emph{same} as for $\alpha$-Fe$_2$O$_3$):

When $\vec{L}$ lies along high-symmetry directions (parallel to the two-fold axis ($\equiv x$), in plane and orthogonal to the two-fold axis ($\equiv y$) and along the high-symmetry axis ($\equiv z$), one obtains the following:

\begin{eqnarray}
\ten{\alpha}^{(3)}_{L\parallel x}&=&\left(\begin{smallmatrix}
 0 & \alpha_{12}& \alpha_{13} \\
\alpha_{21}& 0 & 0 \\
 \alpha_{31} & 0 & 0 \\
\end{smallmatrix}\right)
\nonumber\\
\ten{\alpha}^{(3)}_{L\parallel y}&=& \left(
\begin{smallmatrix}
 \alpha_{11}& 0 & 0 \\
 0 & \alpha_{22} & \alpha_{23} \\
 0 & \alpha_{32} &\alpha_{33} \\
\end{smallmatrix}
\right)
\nonumber\\
\ten{\alpha}^{(3)}_{L\parallel z}&=&\left(
\begin{smallmatrix}
 \alpha_{11} & 0 & 0 \\
 0 & \alpha_{11} & 0 \\
 0 & 0 & \alpha_{33} \\
\end{smallmatrix}
\right)
\end{eqnarray}

One can see that these expressions are identical to the MPG-symmetrised tensors for MPGs $2'/m$ and $2/m'$, including the correct number of independent parameters (see Appendix  \ref{Appendix_generating_tensors}, section \ref{Sec: ME tensors}).
\newline

\subsubsection{Co-rotating component of the magnetoelectric tensor}
\label{Sec: ME TEnsor_Cor} 

As we saw previously, it is convenient to express the linear term-$\vec{L}$ in terms of $3\times 3$ matrices connecting the order parameters with the magnetic indices, so one can immediately spot the presence of a non-traceless component and decompose it into co-rotating and non-co-rotating components.  This is more conveniently done on the \emph{inverse} magneto-electric tensor $\ten{\alpha}^{\mathrm{T}}$.  The inverse tensor connects the induced magnetisation $\vec{M}$ to the applied electric field $\vec{E}$ through the formula $\mu_0 M_i=\alpha^{\rm{T}}_{ij}E_j$ and  has free spin indices.  In the appropriate units, $\alpha^{\rm{T}}_{ij}=\alpha_{ji}$ and the two matrices are the transpose of each other.

 For this, we re-write the polarisation in terms of the canonical matrices in Equation \ref{eq: alter_canonical_3}:
 
 \begin{equation}
\ten{\alpha}^{\mathrm{T}}=\ten{U}\cdot (L_x, L_y,L_z)^{\mathrm T}
\end{equation}

with

\begin{equation}
\ten{U}^{\mathrm{T}}= N^{(A_{2g/u})} E_z+N^{(E_{g/u})}_{(1)} E_x +N^{(E_{g/u})}_{(2)} E_y
\end{equation}

Once again,  the only non-traceless matrix is $N^{(A_{2g})}$ and calling its trace $\tau$,  the co-rotating expression of the magneto-electric tensor becomes:

\begin{equation}
\label{Eq: ME_corotating_version1} 
\ten{\alpha}^{\mathrm{T} cor}=\tau \begin{pmatrix}
0& 0 & L_x\\
0 & 0 & L_y\\
0& 0& L_z
\end{pmatrix}
\end{equation}

and for the direct effect

\begin{equation}
\label{Eq: ME_corotating_version1} 
\ten{\alpha}^{cor}=\tau \begin{pmatrix}
0& 0 & 0\\
0 & 0 & 0\\
L_x& L_y & L_z
\end{pmatrix}
\end{equation}

Note that the non-traceless matrix $N^{(A_{2g})}$ in Equation \ref{eq: alter_canonical_3} can itself be decomposed into a pure-trace part and a traceless component as

\begin{equation}
\begin{pmatrix}
b&0&0\\
0&b&0\\
0&0&c
\end{pmatrix}
=
\frac{2b+c}{3}
\begin{pmatrix}
1&0&0\\
0&1&0\\
0&0&1
\end{pmatrix}
+
\frac{b-c}{3}
\begin{pmatrix}
1&0&0\\
0&1&0\\
0&0&-2
\end{pmatrix}.
\end{equation}

The traceless component transforms with the same \irrep\ $A_{2g}$ but is not co-rotating and is only permitted in the presence of anisotropy.

Once again, one can obtain the same result  symmetrising  directly the $\bm{\mathcal{A}}$ tensor, which in this case is a vector with the expression:

\begin{equation}
\bm{\mathcal{A}}=\begin{pmatrix}
0\\
0 \\
c_3
\end{pmatrix}
\end{equation}

Hence, the co-rotating tensor becomes:

\begin{eqnarray}
\alpha^{cor}_i&=&\mathcal{A}_i \delta_{jk}L_k\nonumber\\
\ten{\alpha}^{cor}&=&c_3\begin{pmatrix}
0& 0 & 0\\
0 & 0 & 0\\
L_x& L_y & L_z
\end{pmatrix}
\end{eqnarray}

which has the same expression as the previous Equation \ref{Eq: ME_corotating_version1}.

\subsubsection{Canonical expression of the inverse magneto-electric tensor}

Finally, I write the expression of the \emph{inverse} magneto-electric tensor in fully canonical form using the canonical vectors in Equation \ref{eq: wmag_canonicals}.  On the mixed-basis polynomials in Equation \ref{eq: E_L} (linear in $\vec{L}$), one can write:

\begin{eqnarray}
\label{eq: ME_inverse_canonical}
\ten{\alpha}^{\rm{T}{(1)}}&=&\sum_{r=1}^2 V_r^{(A_{1g/u})}X_i^{(A_{1u})}
+
\sum_{r=1}^{3}\sum_{\alpha=1}^{2}V^{(E_u)}_{r,(\alpha)}X^{(E_u)}_{r,(\alpha)}\nonumber\\
&=&\left(\begin{smallmatrix}
 b_2 L_z+b_3 L_y&b_3 L_x& b_1 L_x \\
b_3 L_x& b_2 L_z-b_3 L_y & b_1 L_y \\
 a_2 L_x& a_2 L_y & a_1  L_z \\
\end{smallmatrix}\right)
\end{eqnarray} 

which has exactly the same form and number of parameters as Equation \ref{eq: ME_tensor_linear_compact}.

\section{Exchange multiplet and the Landau theory}
\label{Sec: Landau_Theory_connection}

As explained above, the Bertaut--Izyumov framework is closely tied to the Landau theory of phase transitions, which is the main reason it became prevalent among an important section of the magnetism community.  The connection between altermagnetism and the Landau theory has been discussed by McClarty \textit{et al.}, \cite{McClarty2024, Schiff2025} but here it is worth making an explicit connection with exchange multiplets, especially since it helps clarify the link between primary and secondary order parameters.  In the Landau scheme, continuous phase transitions occur when the coefficient of a quadratic term in the free energy changes sign as a function of temperature.  Since quadratic scalar invariants always involve a single \irrep, only one \irrep~mode can condense at a given temperature.  Single-\irrep~quadratic invariants are always isotropic even when the \irrep~is multi-dimensional.  Hence, the linear combination of single \irrep~modes that actually condensed (known as the `direction of the order parameter') is established by higher-order terms in the free energy expansion.

Based on exchange multiplet theory, one can decompose the Landau free energy in a manner that is particularly convenient if the magnetic anisotropy is weak.  In the absence of external fields, one writes

\begin{equation}
\label{eq: Landau_zero field}
F(T)=a(T-T_c) \rho^2+b \rho^4+\dots+ \lambda \Phi_{an}(\eta_i)
\end{equation}

with

\begin{equation}
\rho^2
=\sum_i |\eta_i|^2
\end{equation}

where $\lambda$ is the `small' anisotropy parameter, and the sum runs over all the \irreps~in the multiplet and their individual components $\eta_i$. The two terms represent, respectively, the rotationally invariant (exchange-driven) and anisotropy components of the Landau free energy.  The anisotropy part $\Phi_{an}(\eta_i)$ contains combinations of $\eta_i$ that are not rotationally invariant.  At the lowest order: 

\begin{equation}
\label{eq: non_inv_quadratic}
 \Phi_{an}(\eta_i)=\sum_i \alpha_i|\eta_i|^2+\dots+\sum_{i} \sum_j  \Phi_c (\eta_i , \xi^{(j)}_i)
 \end{equation}

where $\sum_i \alpha_i=0$, the $\xi^{(j)}_i$ are secondary order parameters transforming with the same \irrep~as $\eta_i$ and $\Phi_c (\eta_i , \xi^{(j)}_i)$ are the linear coupling terms between primary and secondary order parameters.  One can easily show that this term leads to a single \irrep~condensing at the transition temperature, as in the general case.\newline

The situation is particularly simple when the generating \irrep~of the \emph{primary} order parameter is 1D and real --- the case that I have been discussing throughout this paper.  In this case, the structure of the exchange multiplet mirrors that of the axial representation and we have:

\begin{eqnarray}
\rho^2&=&L^2\nonumber\\
\Phi_{an}(\eta_i)&=& \Phi_{an}(\vec{L}, \bm{\Lambda}^{(i)})
\end{eqnarray}

where $\bm{\Lambda}^{(i)}$ are secondary magnetic order parameters (see below).


\subsection{Coupling with external fields and additional order parameters}

For weakly anisotropic antiferromagnets in an external magnetic field, one can extend Equation \ref{eq: Landau_zero field} as:

\begin{eqnarray}
\label{eq: Landau_with field}
F(T)&=&F^{cor}(T)+\lambda \tilde{F}(T)\nonumber\\
F^{cor}(T)&=&a(T-T_c) \rho^2+b \rho^4+\dots+\mathcal{A}_iL_j E_iH_j+\dots\nonumber\\
\tilde{F}(T)&=& \Phi_{an}(\vec{L})+ m_j(\vec{L}) H_j+ \tilde{\alpha}_{ij}(\vec{L}) E_iH_j+\dots+ \sum_i \mathcal{I}^{(i)}_{jk} L_j \Lambda^{(i)}_k
\end{eqnarray}

where the term $\mathcal{A}_iL_j$ is the co-rotating part of the magneto-electric tensor (see Section \ref{Sec: ME TEnsor_Cor}) and all the non-co-rotating components are collected in the third line of Equation \ref{eq: Landau_with field}.	This structure is particularly simple because free energy minimisation is performed as a function of $L_x, L_y, L_z$, without any need to refer to specific \irreps~in the exchange multiplet.  Secondary order parameters are connected in the usual way to $\vec{L}$ by a separate minimisation, involving a quadratic term in the $\bm{\Lambda}^{(i)}$ (omitted here for simplicity).

Once again, if the generating \irreps~of both the primary and the secondary order parameters are 1D,  a considerable simplification occurs on the structure of the intertwiners, which is stated here without proof (which would follow the general lines of Equation \ref{eq: covariance_odd}):\newline

\begin{center}
\fbox{%
\parbox{0.9\linewidth}{%
The coupling terms $\mathcal{I}^{(i)}_{jk} L_j \Lambda^{(i)}_k$ are \emph{covariant} iff the intertwiner tensors $\bm{\mathcal{I}}^{(i)}$ are symmetrised by the \emph{product} of $\chi^{\bm{L}}_{bip} \chi^{\bm{\Lambda}^{(i)}}_{bip}$\newline
}}
\end{center}

This statement is extremely important to define the general applicability of the framework I have discussed.  In the presence of secondary magnetic order parameters, the magnetic structure is no longer strictly collinear.  However, the framework is still applicable, provided that the exchange multiplet of the \emph{primary} order parameter is a bipartition \irrep.  This will be shown by way of an example in the next section. 

\subsubsection{Example: secondary order parameters in orthoferrites} 

Orthoferrites are a classic case where canted antiferromagnetism/weak ferromagnetism occur depending on the direction of the primary order parameter \cite{White1969}.  In the conventional $\hmn{Pnma}$ setting (point group $\hmn{mmm}$) and adopting the historical G, C, A, F classification of $\Gamma$-point magnetic ordering in perovskites,  we have the following:

\begin{eqnarray}
\chi^G_{bip}&=&B_{3g}\nonumber\\
\chi^A_{bip}&=&B_{1g}\nonumber\\
\chi^C_{bip}&=&B_{2g}\nonumber\\
\chi^F_{bip}&=&A_g
\end{eqnarray}

The primary order parameter corresponds to G collinear ordering.  The intertwiners are therefore symmetrised with products $\chi^G \chi^A=B_{2g}$,  $\chi^G \chi^C=B_{1g}$ and $\chi^G \chi^F=B_{3g}$.

\begin{eqnarray}
\mathcal{I}^{A}_{ij}&=&\left(
\begin{smallmatrix}
 0 & 0& c_{13} \\
 0& 0 & 0 \\
 c_{31} & 0 & 0 \\
\end{smallmatrix}
\right)\nonumber\\
\mathcal{I}^{C}_{ij}&=&\left(
\begin{smallmatrix}
 0 & c_{12}& 0 \\
 c_{21}& 0 & 0 \\
 0 & 0 & 0 \\
\end{smallmatrix}
\right)\nonumber\\
\mathcal{I}^{F}_{ij}&=&\left(
\begin{smallmatrix}
 0 & 0& 0 \\
 0& 0 & c_{23} \\
 0 & c_{32} & 0 \\
\end{smallmatrix}
\right)
\end{eqnarray}

From these, one can derive the expressions of the secondary order parameters:

\begin{eqnarray}
\label{eq: secondary_coupling_ortho}
\bm{\Lambda}^A&=&\left( c_{13} L_z, 0, c_{31} L_x\right) \nonumber\\
\bm{\Lambda}^C&=&\left( c_{12} L_y, c_{21} L_x, 0 \right) \nonumber\\
\bm{\Lambda}^F&=\vec{m}=&\left( 0, c_{23} L_z, c_{32} L_y \right) 
\end{eqnarray}

where $\bm{\Lambda}^F$ is also the weak magnetisation $\vec{m}$.  Equation \ref{eq: secondary_coupling_ortho} is in full agreement with the well-known MPG analysis of canted structures in orthoferrites, but it is valid for a completely generic direction of the primary order parameter.

\section{Implications for the classification of magnets}

\subsection{Fine-grained classification}

From the discussion so far, the advantage of a granular classification of magnets based on the exchange multiplet of the primary order parameter should be completely clear.  The generating \irrep\ is the natural symmetry label of fine-grained `classes' because it determines both the exchange multiplet and, as shown above, the complete hierarchy of covariant time-reversal-odd tensors and secondary order parameters. Hence, each granular `class of magnets' can simply be denoted by the crystal class of parent group and the point-group bipartition \irrep~of the primary order parameter.  For example, for $\alpha$-Fe$_2$O$_3$ and Cr$_2$O$_3$ the notation could be $\hmn{-3m} (A_{2g})$ and $\hmn{-3m} (A_{2u})$, respectively.  For (quasi)-collinear magnets, this notation is completely equivalent to the spin-group notations $\bar{3}^1m^2$ and $\bar{3}^2m^1$.  For genuinely non-collinear structure with multi-dimensional generating \irreps, the \irrep~notation may actually be more convenient than introducing higher-order spin groups.  For example, for the antiperovskites discussed in Ref. \onlinecite{Radaelli2025ColorSymmetry}, the appropriate class would read  $\hmn{m-3m} (E_{g})$, which is rather compact.  All in all, this analysis vindicates both Izyumov's original intuition and the more recent introduction of spin groups.  Contrary to a common perception, the usefulness of spin groups is not confined to the isotropic exchange limit. The same symmetry framework provides a unified construction of covariant time-reversal-odd tensors in both the isotropic and anisotropic cases, while also identifying the isotropic (co-rotating) component of these tensors whenever it exists. \newline

\subsection{Coarse-grained classification}

Likewise, the exchange multiplet framework also informs the principles of a coarser-grained classification of magnets.  The exchange-multiplet framework naturally preserves the traditional distinction between ferro/ferrimagnets and antiferromagnets. The former are generated by the totally symmetric exchange \irrep~($A_{1g}$ or $A_1)$, whereas the latter comprise all remaining exchange multiplets. From this viewpoint, antiferromagnetism remains the overarching symmetry class, with further subdivisions reflecting the symmetry of the generating \irrep.  For zone-centre magnets, a further distinction can be made between \emph{gerade} generating \irreps~(allowing parity-even, time-reversal-odd responses such as alternating spin textures), \emph{ungerade} generating \irreps~(allowing parity-odd, time-reversal-odd responses such as the magneto-electric effect) and non-centrosymmetric systems (allowing both types of responses).  Within the spintronics literature, it is useful to distinguish altermagnets from other $\Gamma$-point antiferromagnets according to their equilibrium spin splitting. However, this distinction should be regarded as a refinement within the broader class of antiferromagnets rather than as a replacement for it.

\subsection{Beyond quasi-collinear structures}

In truly non-collinear zone-centre magnetic structures, the exchange multiplet generator is a multi-dimensional \irrep.  In general, one writes:

\begin{equation}
\label{eq: non_multiplet}
\Gamma_{mult}=\Gamma_{bip} \otimes A
\end{equation}

where $\Gamma_{bip}$ is two-or three-dimensional.  This immediately invalidates the quasi-collinear form of the intertwiner theorem, since

\begin{equation}
\Gamma_{bip} \otimes \Gamma_{bip} \ne A_{1g}
\end{equation}

Instead, at any intertwiner order, one needs to consider all the components of the $\Gamma_{bip} \otimes \Gamma_{bip}$ product and build different intertwiners with different symmetrisations.  This procedure is not complex in principle, and will be developed in future work.

\subsection{Beyond the zone centre, beyond magnetism}

A significant part of this framework also survives for non-$\Gamma$-point structures.  Exchange multiplets, being completely general, can be defined for any magnetic structure and can be used as the basis for a more general classification.  However, as previously explained, away from the $\Gamma$ point all local intertwiners are zero, since they are averaged over all the parent-group operations, including translations.  It is precisely the phase factors associated with translations that lead to the cancellation of the local intertwiners.  This is obviously true for magnetic structures associated with paramagnetic MPGs (all incommensurate and even-period commensurate), but it remains true for odd-period commensurate, though the latter do not have paramagnetic MPGs and can couple non-locally to a net magnetisation, for example.  Perhaps more interestingly, the framework remains completely applicable to time-reversal-even magnetic order parameters, such as the vector chirality, which may condense at the zone centre.  This observation naturally leads to a Bertaut--Izyumov description of anti-altermagnetism, which will be discussed in a future paper.  More broadly, time reversal is not fundamental to the Bertaut--Izyumov framework or, specifically, to the intertwiner approach developed here. The same construction could therefore be applied, essentially unchanged, to a non-magnetic continuous order parameter transforming under a parent symmetry group, yielding the symmetry-constrained form of tensorial properties that depend covariantly on that order parameter. The recently discovered polar vortices in oxide superlattices may provide a natural testbed for such an extension.\cite{Yadav2016}

\section{Conclusions}

The intertwiner approach to time-reversal-odd tensor calculations based on the Bertaut--Izyumov exchange multiplet framework, which has been developed in this paper, represents a significant departure both from the `traditional' approach based on distinct magnetic point groups for different directions of the order parameter $\vec{L}$ (the N\'eel vector for antiferromagnets) and from more recent approaches that employ spin groups specifically in the context of altermagnetism.  Although spin groups and exchange multiplets are closely related, the fundamental idea that is uniquely enabled by the exchange multiplet framework is that the order parameter is treated as a \emph{continuous variable}, and that time-reversal-odd tensors are expanded in powers of the order parameter via symmetry-adapted polynomials in $\vec{L}$ and tensor intertwiners.  Magnetic-point-group extinction rules for certain tensor components are recovered for all directions through the nodes in the polynomials and the form of the symmetry-adapted intertwiners.  One major advantage of the intertwiner framework is that it factorises the calculation of tensorial properties. The magnetic information enters through the symmetry-adapted projections that determine the appropriate intertwiner forms, which are obtained by straightforward group-theoretical methods. The remainder of the problem, and most of the computational complexity, resides in constructing the canonical polynomials in the order parameter and the physical variables. These depend only on the paramagnetic point group, and can therefore be calculated once and reused for different magnetic structures and tensorial properties. This approach is valid provided that the expansion in the order parameter is \emph{controlled} and the parent (paramagnetic) symmetry remains significant in the magnetic phase, which is expected to be valid for low to moderate values of the spin-lattice interactions. This approach to the calculation of tensorial properties can be implemented rather simply, and appears to be ideally suited both for the current generation of experiments and for density-functional-theory calculations, where the order parameter can be varied continuously, either experimentally or computationally.

\appendix

\section{The Bertaut--Izyumov method}
\label{app: B-I recipe}

Here, I provide a summary of the steps employed within the Bertaut--Izyumov framework to construct candidate magnetic structures given a parent group and a propagation vector. This method is implemented in popular magnetic structure refinement codes such as FULLPROF\cite{Rodriguez-Carvajal1993} and SARAH \cite{Wills2000}, and has been the basis for the solution of the majority of magnetic structures since the 1980s.

\begin{enumerate}
\item Identify the subgroup of the crystal class (rotational part of the space group) that stabilises $\vec{k}$, known as the \emph{little co-group}.
\item Identify the subgroup of the space group (known as the \emph{little group})  with rotational elements corresponding to the little co-group.
\item Check the integrity of the Wyckoff orbit, i.e., that all magnetic sites are still generated by the little group from a single site. If not, the Wyckoff orbits have split and they need to be treated independently.
\item Construct the reducible representation of the little group (known as \emph{magnetic representation}), obtained as a product of the $\vec{k}$-dependent \emph{permutation representation} of the Wyckoff orbit times the axial vector representation.  This representation acts on the linear space of all possible sets of complex vectors, each associated with a magnetic site $j$ on the Wyckoff orbit and having the correct phases upon pure translations.  More details about the permutation representation are discussed in Appendix \ref{Appendix: Perm_rep}.
\item Select an \irrep~(say, $\Delta$) contained within the \emph{magnetic representation}, and project the one-atom basis vectors onto that \irrep, so as to obtain an appropriate basis set of complex vectors $\vec{S}^\Delta_{\alpha} (j)$ associated with orbit sites.
\item The magnetic structure, i.e., a collection of real axial vectors (magnetic moments) on each site $j$ for all unit cells $n$,  is then:

\begin{equation}
\vec{m}(n,j) =\sum_\alpha c_\alpha \vec{S}^\Delta_\alpha (j) e^{-2 \pi i \vec{k} \cdot \vec{R}_n}+ c.c.
\end{equation} 

where the $\vec{S}^{\Delta}_{\alpha}(j)$ are complex basis vectors associated with magnetic site $j$ and the selected \irrep~$\Delta(\vec{k})$. The index $\alpha$ runs over the dimension and multiplicity of the \irrep, while the coefficients $c_\alpha$ specify the particular magnetic structure within that \irrep.  Note that each magnetic structure of this kind is associated with both $\vec{k}$ and $-\vec{k}$, which are generally inequivalent.
\end{enumerate}

The procedure can be extended to multi-$\vec{k}$ structures where different $\vec{k}$s in the star represent different components rather than domains, or to magnetic structures described by multiple \irreps.  

\section{The permutation representation}
\label{Appendix: Perm_rep}

The $\vec{k}$-dependent permutation representation $\Gamma_{perm,\vec{k}}$ is a fundamental concept in magnetic structure analysis and is employed by the most popular analysis codes (FULLPROF\cite{Rodriguez-Carvajal1993}, SARAH \cite{Wills2000}) to construct magnetic representations. Let $G$ be a space group and let $X=\{x_j\}$ be a Wyckoff orbit, i.e., a set of crystallographically equivalent sites within the unit cell at the origin, generated from a single site by the action of the space group, with all sites referred back to the reference unit cell.  The permutation representation acts on a vector space spanned by basis vectors $\{e_j\}$, one for each site in  $\{X\}$, which is  equivalent sites in all unit cells through the Bloch condition:

\begin{equation}
e_{j, \vec{R}}=e_{j, 0} \,e^{2 \pi \vec{k} \cdot \vec{R}}
\end{equation}

 so that the $\{e_j\}$ denotes a basis of Bloch states at wave vector $\vec{k}$.  The $\{e_{j,0}\}$ can be thought of as arrays where the $j$ `slot' contains $1$, while all the other slots contain $0$.  

For each $g\in G$, the action on the orbit is defined by
\[
g x_j = x_{j'} + \mathbf{T}_j(g),
\]
where $x_{j'}\in X$ and $\mathbf{T}_j(g)$ is the pure lattice translation required to express the transformed site in terms of the chosen representative $x_{j'}$ of the orbit.  
The associated $\mathbf{k}$-dependent representation is then
\[
\mathcal D_{\mathbf k}(g)e_j
=
e^{-2 \pi i\mathbf k\cdot \mathbf{T}_j(g)} e_{j'} ,
\]
where the sign convention in the exponential depends on the Fourier-transform convention adopted.

Equivalently, in matrix form,
\[
\left[\mathcal D_{\mathbf k}(g)\right]_{j'j}
=
e^{-2 \pi i\mathbf k\cdot \mathbf{T}_j(g)},
\]
with all other matrix elements equal to zero.

With this definition, the magnetic representation $\Gamma_{mag}$ becomes $\Gamma_{mag}=\Gamma_{perm, \vec{k}} \otimes \Gamma_{axial}$, where $\Gamma_{axial}$ is the axial vector representation. This is implemented, for example, in the program BASIREP, which is part of the FULLPROF suite,\cite{RodriguezCarvajal2025FullProf} and can be used to construct all possible \irreps\ associated with a given magnetic representation.

\section{Generating Shubnikov groups from 1D real \irreps}
\label{Appendix: bicolor_irrep}

Spin groups, magnetic point groups and magnetic space groups are all examples of bi-colour groups, and are in a one-to-one correspondence with the celebrated Shubnikov groups.  The most direct connection between bi-colour groups and their characteristic representation (the colour permutation representation) is provided by the so-called graph construction.  Given a group $G$, a 1D real \irrep~$\chi$ of $G$  (a sign character), $\chi: G\mapsto \{\pm1\}$, is just a group homomorphism into $\mathbb{Z}_2$.  The \emph{graph} of this homomorphism

\begin{equation}
\Gamma_\chi = \{(g,\chi(g))\} \subset G \times \mathbb{Z}_2
\end{equation}

is a bi-colour group that is generated by the representation.   Type-I (white) Shubnikov groups are generated by the trivial representation.  Type-III (black and white, zone centre) and Type-IV (black and white, zone boundary)  Shubnikov groups are generated by non-trivial representations.  By contrast, Type-II (grey or paramagnetic) groups are the full direct product groups $G \times \mathbb{Z}_2$ and are not generated via a representation.  Hence, in all examples above but the paramagnetic case, we can use the representation and group language interchangeably: we would say that the group in question is the \emph{graph} of the representation, and that the representation is a $\mathbb{Z}_2$-\emph{grading} (or \emph{grading} for short) of the group.  Conversely, if $H$ is a subgroup of $G$ of index 2, which transforms sites with spin up into other sites with spin up, the totally symmetric representation of $H$ induces a representation on $G$ (the colour permutation representation), which decomposes in the totally symmetric representation plus a second real 1D representation --- the bipartition representation (see below for the general derivation).

\section{Collinear antiferromagnetic structures at the $\Gamma$ point}

 Collinear antiferromagnetic $\Gamma$-point structures on a single Wyckoff orbit are described by a bipartition pattern, i.e. a 1D real sign character contained in the permutation representation of the orbit. Let $\left\{x\right\}$ be a magnetic orbit generated by a space group $G$ from a single position $x_0$, and let $H^\prime$, the stabiliser of one colour class, be a subgroup of index 2 in $G$ containing the site stabiliser of $x_0$. The coset decomposition of $G$ is
 
 \begin{equation}
G=H^\prime\cup gH^\prime
\end{equation}

while the Wyckoff orbit set decomposes into two orbits as

\begin{equation}
\{x\}=H^\prime x_0\cup gH^\prime x_0.
\end{equation}

The induced representation ${\rm \operatorname{Ind}}_{H^\prime}^G(1)$ is the colour permutation representation induced by the totally symmetric representation of $H^\prime$. Its decomposition contains two irreducible components: the totally symmetric representation and a second 1D \irrep, the bipartition irrep $\chi_{bip}$. A physical tensor is then built as a tensor product $X(L)\otimes B$ of a spin-dependent factor $X(L)$, transforming as an appropriate tensor power of the axial-vector representation, and a bipartition factor $B$ transforming as $\chi_{bip}$. On restriction to the subgroup that stabilises L to within a 1D real character $\chi_{stab}$, the total tensor transforms as $\chi_{stab}\otimes\chi_{bip}$, and the corresponding magnetic point group is obtained by priming those operations with character -1. The corresponding spin group is obtained analogously from $\chi_{bip}$.

\section{Simple proofs}
\label{Appendix: simple_proofs}
 
\subsection{Covariant tensors are `physical materials tensors' in the ordinary MPG sense}

As already discussed, the MPG is obtained from the stabiliser subgroup $H$ of a particular direction of $\vec{L}$, together with the associated sign character. Let $g \in H$: then, for that particular direction of $\vec{L}$, $g \vec{L}=\chi_{MPG}(g)  \vec{L}$ with $\chi_{MPG}(g) = \pm1$ and

\begin{equation}
\alpha'_{ij} = \alpha_{ij}(\vec{L}') = \chi_{MPG}(g) \alpha_{ij} 
\end{equation}

since $\chi_{MPG}(g)^{2n+1} = \chi_{MPG} (g)$ for all odd powers. This is precisely the condition for a time-reversal-odd tensor
to be MPG-invariant. The same follows trivially for any time-reversal-even tensor, since $\chi_{MPG}(g)^{2n} = +1$ for every
even power.

The inverse statement `all physical materials tensors' in the ordinary MPG sense are covariant' is not obvious.  The physical significance of any non-covariant materials tensors has, to my knowledge, not been discussed in the literature.

\subsection{Intertwiner theorem 1}

Taking the magneto-electric tensor $\alpha_{ij}(\vec{L})$ as an example,  the Landau free energy of certain antiferromagnets contains terms of the kind:

\begin{equation}
F(\vec{L})= (\mathcal{I}_{ijk} L_k +\mathcal{I}_{ijklm} L_kL_m L_n + \dots) E_iH_j
\end{equation}

where 

\begin{equation}
\label{eq: rank2_expansion}
\alpha_{ij}(\vec{L})=\mathcal{I}^{(1)}_{ijk} L_k + \mathcal{I}^{(3)}_{ijklm} L_kL_m L_n + \dots
\end{equation}

is the Landau--Lifshitz expansion of the tensor $\alpha_{ij}(\vec{L})$.  Invariance of the Landau free energy by any symmetry operator of the paramagnetic group requires 

\begin{equation} 
\label{eq: equivariance}
\alpha'_{ij}(\vec{L})=\alpha_{ij}(\vec{L'})
\end{equation}

to be valid to any order of the expansion, with eq. \ref{eq: equivariance} expressing the condition for $\alpha$ and $\vec{L}$ to be \emph{equivariant} (or alternatively, for $\alpha$ to be covariant with $\vec{L}$). One can immediately see that covariant tensors are `physical' in the ordinary MPG sense (see Appendix \ref{Appendix: simple_proofs}). 
\newline

Note that both the covariance requirement and the MPG invariance of covariant tensors are completely general, regardless of the specific \irrep~or sum of \irreps~ defining the transformation properties of $\vec{L}$.  However, something quite remarkable happens for exchange multiplets defining collinear structures, since, as we have seen in eq. \ref{eq: collinear_multiplet}, the generating \irrep~$\chi_{bip}$ of the exchange multiplet is 1D.  Let us consider a term in the expansion in eq. \ref{eq: rank2_expansion}:

\begin{equation}
\alpha^{(3)}_{ij}(\vec{L})= \mathcal{I}^{(3)}_{ijklm} L_kL_m L_n 
\end{equation}

The covariance condition is written as:


\begin{equation}
D_{i i'}D_{j j'} \mathcal{I}^{(3)}_{i'j'klm}  =\mathcal{I}^{(3)}_{ijk'l'm'}  \chi_{bip}^3 D_{k' k} D_{l' l} D_{m' m} 
\end{equation}

Where the $D$'s are rotation matrices associated with the group element $g$.  Multiplying by the inverse matrices and re-assigning the indices one obtains:

\begin{equation}
\label{eq: covariance_odd}
D_{i i'}D_{j j'}D_{k k'}D_{l l'} D_{m m'} \mathcal{I}^{(3)}_{i'j'k'l'm'}  =\chi_{bip}\, \mathcal{I}^{(3)}_{ijklm}   
\end{equation}

where we have used the fact that, for a sign character, $\chi_{bip}^{2n+1}=\chi_{bip}$ for any odd integer.  For even powers, a parallel derivation leads, for example,  to the relation

\begin{equation}
D_{i i'}D_{j j'}D_{k k'}D_{l l'} D_{m m'} D_{n n'} \mathcal{I}^{(4)}_{i'j'k'l'm'n'}  = \mathcal{I}^{(4)}_{ijklmn}   
\end{equation}

 where we have used the fact that, for a sign character, $\chi_{bip}^{2n}=1$ for any even integer. We can conclude that $\mathcal{I}^{(4)}_{ijklmn}$ is \emph{invariant} under operations of the parent group.  We can summarise these findings as follows:
 \newline

\textit{All intertwiners in the Landau--Lifshitz expansion of covariant tensors in powers of the magnetic order parameters transform like the bipartition \irrep~ (for odd powers) or the totally symmetric \irrep~ (for even powers).}

This statement leads to a great simplification in the construction of intertwiners in the case of $\Gamma$-point collinear structures.  In the case of \emph{non-collinear} $\Gamma$-point structures, a very similar statement holds for the linear intertwiner, but the derivation for higher-order intertwiners is somewhat more complicated.

\section{Adapting MTENSOR to generate Cartesian intertwiners}
\label{App: MTENSOR_Adaptation}

Here, I will present a simple method to generate fully Cartesian intertwiners using the program MTENSOR in the Bilbao crystallographic server \cite{PerezMato2015, Gallego2019}.  Taken at face value, MTENSOR is designed to construct tensors that are invariant under a given MPG.  However, `under the bonnet', what MTENSOR does is to symmetrise a tensor on a 1D real \irrep~that corresponds to the MPG, so it contains all the ingredients necessary to symmetrise intertwiners provided that the \irrep~ is replaced by $\chi_{bip}$.  This is done by the following simple steps.

\begin{itemize}
\item Starting from the Jahn symbol of the intertwiner (e.g., $V2[V4]$ for the intertwiner at linear order in $\vec{L}$ of the quadratic piezomagnetic (or altermagnetic) tensor), add an `$a$' in front of the Jahn symbol.  This \textbf{does not} indicate time reversal (as it is usually the case), but rather the fact that the tensor needs to be projected on a non-trivial 1D \irrep.

\item Starting from $\chi_{bip}$, one constructs an `artificial' MPG by priming all the group elements $g$ of the \emph{parent} group $G$ that are associated with $\chi_{bip}(g)=-1$.  Note that this is the same thing as converting (again, artificially) a SG into an MPG by replacing any $g^2$ with $g'$.

\item Finally, construct the appropriate form of the modified intertwiner symbol ($aV2[V4]$ in the example $V2[V4]$  above), symmetrised over the `artificial' MPG.  This is usually done with the custom option in MTENSOR.
\end{itemize}

\section{Canonical $3 \times 36$ forms for parent group $\bar{3}m$}
\label{app: 3times36 matrices}

The matrices here below are partitioned into six $3\times6$ blocks corresponding to the
second Voigt index,

\begin{equation}
R^{(\Gamma)}
=
\bigl(
R^{(\Gamma)}_{11},
R^{(\Gamma)}_{22},
R^{(\Gamma)}_{33},
R^{(\Gamma)}_{23},
R^{(\Gamma)}_{13},
R^{(\Gamma)}_{12}
\bigr).
\end{equation}

\begin{align}
R^{(A_{1g/u})}_{(11)}
&=
\begin{pmatrix}
0 & 0 & 0 & 0 & a & b\\
\dfrac{b}{2}+\dfrac{3c}{2} &
\dfrac{b}{2}-\dfrac{c}{2} &
e &
\dfrac{a}{3} &
0 &
0\\
f &
\dfrac{f}{3} &
g &
h &
0 &
0
\end{pmatrix},
\nonumber\\[2ex]
R^{(A_{1g/u})}_{(22)}
&=
\begin{pmatrix}
0 & 0 & 0 & 0 & \dfrac{a}{3} & c\\
\dfrac{b}{2}-\dfrac{c}{2} &
-\dfrac{3b}{2}-\dfrac{c}{2} &
-e &
a &
0 &
0\\
\dfrac{f}{3} &
f &
g &
-h &
0 &
0
\end{pmatrix},
\nonumber\\[2ex]
R^{(A_{1g/u})}_{(33)}
&=
\begin{pmatrix}
0 & 0 & 0 & 0 & d & e\\
e &
-e &
0 &
d &
0 &
0\\
g &
g &
i &
0 &
0 &
0
\end{pmatrix},
\nonumber\\[2ex]
R^{(A_{1g/u})}_{(23)}
&=
\begin{pmatrix}
0 & 0 & 0 & 0 & e & \dfrac{a}{3}\\
\dfrac{a}{3} &
a &
d &
-e &
0 &
0\\
h &
-h &
0 &
g &
0 &
0
\end{pmatrix},
\nonumber\\[2ex]
R^{(A_{1g/u})}_{(13)}
&=
\begin{pmatrix}
a &
\dfrac{a}{3} &
d &
e &
0 &
0\\
0 &
0 &
0 &
0 &
e &
\dfrac{a}{3}\\
0 &
0 &
0 &
0 &
g &
h
\end{pmatrix},
\nonumber\\[2ex]
R^{(A_{1g/u})}_{(12)}
&=
\begin{pmatrix}
b &
c &
e &
\dfrac{a}{3} &
0 &
0\\
0 &
0 &
0 &
0 &
\dfrac{a}{3} &
\dfrac{b}{2}-\dfrac{c}{2}\\
0 &
0 &
0 &
0 &
h &
\dfrac{f}{3}
\end{pmatrix}
\end{align}

\begin{align}
R^{(A_{2g/u})}_{(11)}
&=
\begin{pmatrix}
j & k & l & m & 0 & 0\\
0 & 0 & 0 & 0 & -3m & -\dfrac{j}{2}-\dfrac{3k}{2}\\
0 & 0 & 0 & 0 & o & 0
\end{pmatrix},
\nonumber\\[2ex]
R^{(A_{2g/u})}_{(22)}
&=
\begin{pmatrix}
k & -j-2k & -l & 3m & 0 & 0\\
0 & 0 & 0 & 0 & -m & \dfrac{k}{2}-\dfrac{j}{2}\\
0 & 0 & 0 & 0 & -o & 0
\end{pmatrix},
\nonumber\\[2ex]
R^{(A_{2g/u})}_{(33)}
&=
\begin{pmatrix}
l & -l & 0 & n & 0 & 0\\
0 & 0 & 0 & 0 & -n & -l\\
0 & 0 & 0 & 0 & 0 & 0
\end{pmatrix},
\nonumber\\[2ex]
R^{(A_{2g/u})}_{(23)}
&=
\begin{pmatrix}
m & 3m & n & -l & 0 & 0\\
0 & 0 & 0 & 0 & -l & -m\\
0 & 0 & 0 & -o & o & -o
\end{pmatrix},
\nonumber\\[2ex]
R^{(A_{2g/u})}_{(13)}
&=
\begin{pmatrix}
0 & 0 & 0 & 0 & l & m\\
-3m & -m & -n & -l & 0 & 0\\
0 & 0 & 0 & 0 & 0 & -o
\end{pmatrix},
\nonumber\\[2ex]
R^{(A_{2g/u})}_{(12)}
&=
\begin{pmatrix}
0 & 0 & 0 & 0 & m & k\\
-\dfrac{j}{2}-\dfrac{3k}{2} &
\dfrac{k}{2}-\dfrac{j}{2} &
-l &
-m &
0 &
0\\
0 & 0 & -o & 0 & 0 & 0
\end{pmatrix}.
\end{align}

\begin{align}
R^{(E_{g/u})}_{(1);(11)}
&=
\begin{pmatrix}
0 & 0 & 0 & 0 & p & q \\
v & -\frac{q}{3}+s+\frac{v}{3} & w & x & 0 & 0 \\
z & a_1 & b_1 & c_1 & 0 & 0
\end{pmatrix},
\nonumber\\[2ex]
R^{(E_{g/u})}_{(1);(22)}
&=
\begin{pmatrix}
0 & 0 & 0 & 0 & r & s \\
-\frac{q}{3}+s+\frac{v}{3} & 2q+2s+v & 2u+w & -p-r-x & 0 & 0 \\
a_1 & -2a_1-z & -b_1 & 3c_1 & 0 & 0
\end{pmatrix},
\nonumber\\[2ex]
R^{(E_{g/u})}_{(1);(33)}
&=
\begin{pmatrix}
0 & 0 & 0 & 0 & t & u \\
w & 2u+w & y & -t & 0 & 0 \\
b_1 & -b_1 & 0 & d_1 & 0 & 0
\end{pmatrix},
\nonumber\\[2ex]
R^{(E_{g/u})}_{(1);(23)}
&=
\begin{pmatrix}
0 & 0 & 0 & 0 & u & r \\
x & -p-r-x & -t & 2u+w & 0 & 0 \\
c_1 & 3c_1 & d_1 & -b_1 & 0 & 0
\end{pmatrix},
\nonumber\\[2ex]
R^{(E_{g/u})}_{(1);(13)}
&=
\begin{pmatrix}
p & r & t & u & 0 & 0 \\
0 & 0 & 0 & 0 & w & x \\
0 & 0 & 0 & 0 & b_1 & c_1
\end{pmatrix},
\nonumber\\[2ex]
R^{(E_{g/u})}_{(1);(12)}
&=
\begin{pmatrix}
q & s & u & r & 0 & 0 \\
0 & 0 & 0 & 0 & x & -\frac{q}{3}+s+\frac{v}{3} \\
0 & 0 & 0 & 0 & c_1 & a_1
\end{pmatrix}.
\end{align}

\begin{align}
R^{(E_{g/u})}_{(2);(11)}
&=
\begin{pmatrix}
-q-3s-v &
-\dfrac{2q}{3}-\dfrac{v}{3} &
-2u-w &
-\dfrac{p+r}{2}-x &
0 &
0\\
0&0&0&0&
-\dfrac{p+3r}{2}&
-q\\
0&0&0&0&
-3c_{1}&
-\dfrac{3a_{1}+z}{2}
\end{pmatrix},
\nonumber\\[2ex]
R^{(E_{g/u})}_{(2);(22)}
&=
\begin{pmatrix}
-\dfrac{2q}{3}-\dfrac{v}{3} &
q-s-v &
-w &
-\dfrac{p+r}{2}+x &
0 &
0\\
0&0&0&0&
\dfrac{r-p}{2}&
-s\\
0&0&0&0&
-c_{1}&
\dfrac{a_{1}-z}{2}
\end{pmatrix},
\nonumber\\[2ex]
R^{(E_{g/u})}_{(2);(33)}
&=
\begin{pmatrix}
-2u-w &
-w &
-y &
-t &
0 &
0\\
0&0&0&0&
-t&
-u\\
0&0&0&0&
-d_{1}&
-b_{1}
\end{pmatrix},
\nonumber\\[2ex]
R^{(E_{g/u})}_{(2);(23)}
&=
\begin{pmatrix}
-\dfrac{p+r}{2}-x &
-\dfrac{p+r}{2}+x &
-t &
-w &
0 &
0\\
0&0&0&0&
-u&
\dfrac{r-p}{2}\\
0&0&0&0&
-b_{1}&
-c_{1}
\end{pmatrix},
\nonumber\\[2ex]
R^{(E_{g/u})}_{(2);(13)}
&=
\begin{pmatrix}
0&0&0&0&
-2u-w&
-\dfrac{p+r}{2}-x\\
-\dfrac{p+3r}{2}&
\dfrac{r-p}{2}&
-t&
-u&
0&
0\\
-\dfrac{3a_{1}+z}{2}&
\dfrac{a_{1}-z}{2}&
-b_{1}&
-c_{1}&
0&
0
\end{pmatrix},
\nonumber\\[2ex]
R^{(E_{g/u})}_{(2);(12)}
&=
\begin{pmatrix}
0&0&0&0&
-\dfrac{p+r}{2}-x&
-\dfrac{2q}{3}-\dfrac{v}{3}\\
0&0&0&0&
-q&
-s\\
0&0&0&0&
-\dfrac{3a_{1}+z}{2}&
\dfrac{a_{1}-z}{2}
\end{pmatrix}.
\end{align}

\section{Examples of the correspondence between fully Cartesian and canonical intertwiners}
\label{Appendix_generating_tensors}

In this section, I present two examples of Cartesian intertwiners obtained using the program MTENSOR as discussed in Appendix \ref{App: MTENSOR_Adaptation} and the correspondence between their coefficients and those of the canonical forms.  These forms are the same for tensors of opposite parities and symmetrised with the opposite-parity \irreps, as indicated in the captions.  For example, the intertwiner for the piezomagnetic tensor symmetrised with a \textit{gerade} \irrep\ is the same as the  intertwiner for the piezotoroidic tensor symmetrised with an \textit{ungerade} \irrep.

\subsection{Intertwiners for the linear piezomagnetic/$k^2$-altermagnetic tensor}

The linear piezomagnetic tensor and the altermagnetic tensor (quadratic in $\vec{k}$) have Jahn symbols $aeV[V^2]$.  The corresponding intertwiners have Jahn symbols $V^2[V^2]$ (linear in $\vec{L}$, 8 independent parameters) and $V[V^2][V^3]$ (cubic in $\vec{L}$, 28 independent parameters).  The Cartesian expression for the former is reported in Table \ref{Tab: piezo_linear_L}, while the latter is too large to fit in print, but can be easily re-generated using the method discussed in Appendix \ref{App: MTENSOR_Adaptation}.

\begin{table}[h]
\centering
\renewcommand{\arraystretch}{1.2}
\begin{tabular}{c|cccccc}
$c_{ijk}$ & 1 & 2 & 3 & 4 & 5 & 6 \\
\hline
11 & 0 & 0 & 0 & 0 & $c_{115}$ & $c_{116}$ \\
21 & $-2c_{116}-c_{121}$ & $-c_{121}$ & $c_{123}$ & $-c_{115}$ & 0 & 0 \\
31 & $c_{311}$ & $-c_{311}$ & 0 & $c_{314}$ & 0 & 0 \\
12 & $c_{121}$ & $2c_{116}+c_{121}$ & $c_{123}$ & $-c_{115}$ & 0 & 0 \\
22 & 0 & 0 & 0 & 0 & $-c_{115}$ & $-c_{116}$ \\
32 & 0 & 0 & 0 & 0 & $-c_{314}$ & $-c_{311}$ \\
13 & $c_{131}$ & $-c_{131}$ & 0 & $c_{134}$ & 0 & 0 \\
23 & 0 & 0 & 0 & 0 & $-c_{134}$ & $-c_{131}$ \\
33 & 0 & 0 & 0 & 0 & 0 & 0
\end{tabular}
\caption{\label{Tab: piezo_linear_L} Cartesian intertwiner with Jahn symbols $V^2[V^2]$ and $eV^2[V^2]$, symmetrised with \irreps~$A_{2g}$ and $A_{2u}$, respectively}
\end{table}

I will demonstrate that, when converted to a polynomial basis in $\vec{L}$, these tensors are expressed precisely in terms of the $M$ canonical matrices from Equation \ref{Eq: Voigt_canonical}.   \newline

Converting the $V^2[V^2]$ intertwiner into the canonical form (canonical basis in $\vec{L}$) one obtains:\newline

\noindent\textbf{\boldmath$A_{2g} \equiv L_z$}
\begin{align}
\begin{pmatrix}
A & -A & 0 & B & 0 & 0 \\
0 & 0 & 0 & 0 & -B & -A \\
0 & 0 & 0 & 0 & 0 & 0
\end{pmatrix}.
\end{align}

\noindent\textbf{\boldmath$E_g\equiv (L_x, L_y)$}
\begin{align}
\begin{pmatrix}
0 & 0 & 0 & 0 & C & D \\
E & F & G & -C & 0 & 0 \\
H & -H & 0 & I & 0 & 0
\end{pmatrix}
\qquad
&
\begin{pmatrix}
-F & -E & -G & -C & 0 & 0 \\
0 & 0 & 0 & 0 & -C & -D \\
0 & 0 & 0 & 0 & -I & -H
\end{pmatrix}.
\end{align}

with 

\begin{align}
A &= c_{131}, &
B &= c_{134}, \notag\\
C &= c_{115}, &
D &= c_{116}, \notag\\
E &= -2c_{116}-c_{121}, &
F &= -c_{121}, \notag\\
G &= -c_{123}, &
H &= c_{311}, \notag\\
I &= c_{314}.
\end{align}

Note that there is a total of 8 independent parameters, since: 

\begin{eqnarray}
D&=&\frac{F-E}{2},
\end{eqnarray}

precisely as in the canonical forms in Equation \ref{Eq: Voigt_canonical}.

Likewise, converting the $V[V^2][V^3]$ intertwiner into the canonical form (canonical basis in $L^3$) one obtains for the three \irreps:

\noindent\textbf{$A_{1g}$}
\begin{align}
\begin{pmatrix}
0 & 0 & 0 & 0 & A & B \\
B & -B & 0 & A & 0 & 0 \\
C & C & D & 0 & 0 & 0
\end{pmatrix}.
\end{align}

\noindent\textbf{$A_{2g}$}
\begin{align}
\begin{pmatrix}
E & -E & 0 & F & 0 & 0 \\
0 & 0 & 0 & 0 & -F & -E \\
0 & 0 & 0 & 0 & 0 & 0
\end{pmatrix},
\notag\\[3mm]
\begin{pmatrix}
G & -G & 0 & H & 0 & 0 \\
0 & 0 & 0 & 0 & -H & -G \\
0 & 0 & 0 & 0 & 0 & 0
\end{pmatrix},
\notag\\[3mm]
\begin{pmatrix}
I & -I & 0 & J & 0 & 0 \\
0 & 0 & 0 & 0 & -J & -I \\
0 & 0 & 0 & 0 & 0 & 0
\end{pmatrix}.
\end{align}

\noindent\textbf{$E_g$}
\begin{align}
\begin{pmatrix}
0 & 0 & 0 & 0 & K & L \\
M & N & O & -K & 0 & 0 \\
P & -P & 0 & Q & 0 & 0
\end{pmatrix}
\qquad
&
\begin{pmatrix}
-N & -M & -O & -K & 0 & 0 \\
0 & 0 & 0 & 0 & -K & -L \\
0 & 0 & 0 & 0 & -Q & -P
\end{pmatrix},
\notag\\[3mm]
\begin{pmatrix}
0 & 0 & 0 & 0 & R & S \\
T & U & V & -R & 0 & 0 \\
W & -W & 0 & X & 0 & 0
\end{pmatrix}
\qquad
&
\begin{pmatrix}
-U & -T & -V & -R & 0 & 0 \\
0 & 0 & 0 & 0 & -R & -S \\
0 & 0 & 0 & 0 & -X & -W
\end{pmatrix},
\notag\\[3mm]
\begin{pmatrix}
0 & 0 & 0 & 0 & Y & Z \\
A_1 & B_1 & C_1 & -Y & 0 & 0 \\
D_1 & -D_1 & 0 & E_1 & 0 & 0
\end{pmatrix}
\qquad
&
\begin{pmatrix}
-B_1 & -A_1 & -C_1 & -Y & 0 & 0 \\
0 & 0 & 0 & 0 & -Y & -Z \\
0 & 0 & 0 & 0 & -E_1 & -D_1
\end{pmatrix}.
\end{align}

\begin{align}
A &= \frac{1}{4}c_{1511}-\frac{3}{4}c_{1521}, &
B &= \frac{1}{4}c_{1611}-\frac{3}{4}c_{1621}, \notag\\
C &= \frac{1}{4}c_{3111}-\frac{3}{4}c_{3121}, &
D &= c_{3311}, \notag\\[2mm]
E &= \frac{3}{4}c_{1112}-\frac{1}{4}c_{1122}, &
F &= c_{1412}+\frac{1}{4}c_{1511}+\frac{1}{4}c_{1521}, \notag\\
G &= \frac{3}{4}c_{1113}+\frac{3}{4}c_{1123}
     +\frac{1}{2}c_{1133}, &
H &= \frac{3}{2}c_{1413}+\frac{1}{2}c_{1433}
     +\frac{3}{2}c_{1541}, \notag\\
I &= \frac{3}{4}c_{1113}+\frac{3}{4}c_{1123}
     -\frac{1}{2}c_{1133}, &
J &= \frac{3}{2}c_{1413}-\frac{1}{2}c_{1433}
     +\frac{3}{2}c_{1541}, \notag\\[2mm]
K &= \frac{3}{8}c_{1511}+\frac{3}{8}c_{1521}
     +\frac{3}{2}c_{1531}, &
L &= \frac{3}{8}c_{1611}+\frac{3}{8}c_{1621}
     +\frac{3}{2}c_{1631}, \notag\\
M &= -\frac{3}{8}c_{1112}-\frac{3}{8}c_{1122}
     -\frac{3}{2}c_{1132}
     -\frac{3}{4}c_{1611}-\frac{3}{4}c_{1621}
     -3c_{1631}, \notag\\
N &= -\frac{3}{8}c_{1112}-\frac{3}{8}c_{1122}
     -\frac{3}{2}c_{1132}, &
O &= -\frac{3}{2}c_{1312}-\frac{3}{2}c_{1332}, \notag\\
P &= \frac{3}{8}c_{3111}+\frac{3}{8}c_{3121}
     +\frac{3}{2}c_{3131}, &
Q &= -\frac{3}{2}c_{3512}-\frac{3}{2}c_{3532}, \notag\\[2mm]
R &= \frac{3}{8}c_{1511}+\frac{3}{8}c_{1521}
     -\frac{3}{2}c_{1531}, &
S &= \frac{3}{8}c_{1611}+\frac{3}{8}c_{1621}
     -\frac{3}{2}c_{1631}, \notag\\
T &= -\frac{3}{8}c_{1112}-\frac{3}{8}c_{1122}
     +\frac{3}{2}c_{1132}
     -\frac{3}{4}c_{1611}-\frac{3}{4}c_{1621}
     +3c_{1631}, \notag\\
U &= -\frac{3}{8}c_{1112}-\frac{3}{8}c_{1122}
     +\frac{3}{2}c_{1132}, &
V &= -\frac{3}{2}c_{1312}+\frac{3}{2}c_{1332}, \notag\\
W &= \frac{3}{8}c_{3111}+\frac{3}{8}c_{3121}
     -\frac{3}{2}c_{3131}, &
X &= -\frac{3}{2}c_{3512}+\frac{3}{2}c_{3532}, \notag\\[2mm]
Y &= 3c_{1541}, &
Z &= 3c_{1641}, \notag\\
A_1 &= -\frac{3}{2}c_{1113}+\frac{3}{2}c_{1123}
       -6c_{1641}, &
B_1 &= -\frac{3}{2}c_{1113}+\frac{3}{2}c_{1123}, \notag\\
C_1 &= -3c_{1313}, &
D_1 &= 3c_{3141}, \notag\\
E_1 &= -3c_{3513}.
\end{align}

Note that there is a total of 28 independent parameters, since: 

\begin{eqnarray}
L&=&\frac{N-M}{2}\nonumber\\
S&=&\frac{U-T}{2}\nonumber\\
Z&=&\frac{B_1-A_1}{2},
\end{eqnarray}

precisely as in the canonical forms in Equation \ref{Eq: Voigt_canonical}.

\subsection{Intertwiners for the magneto-electric tensor}

The linear magnetoelectric tensor has Jahn symbols $aeV^2$.  The corresponding intertwiners have Jahn symbols $V^3$ (linear in $\vec{L}$, 5 independent parameters) and $V^2[V^3]$ (cubic in $\vec{L}$, 16 independent parameters).  The Cartesian expressions are reported in Tables \ref{Tab: Gen_tensor_3} and  \ref{Tab: Gen_tensor_23}.

Converting the $V^2[V^3]$ intertwiner into the canonical form (canonical basis in $\vec{L}$) one obtains for the three \irreps:

\noindent\textbf{\boldmath$A_{2g}\equiv L_z$}
\begin{align}
\begin{pmatrix}
A & 0 & 0 \\
0 & A & 0 \\
0 & 0 & B
\end{pmatrix},
\end{align}

\noindent\textbf{\boldmath$E_g\equiv (L_x, L_y)$}
\begin{align}
\begin{pmatrix}
0 & C & D \\
C & 0 & 0 \\
E & 0 & 0
\end{pmatrix}
\qquad
&
\begin{pmatrix}
C & 0 & 0 \\
0 & -C & D \\
0 & E& 0
\end{pmatrix},
\end{align}

with

\begin{align}
A &= c_{311}, &
B &= c_{333}, \notag\\
C &= c_{112}, &
D &= c_{113}, \notag\\
E &=c_{131}&
\end{align}

\begin{table}[h]
\caption{\label{Tab: Gen_tensor_3} Cartesian intertwiner with Jahn symbols $V^3$  (magneto-electric tensor) and $eV^3$ (electrotoroidic tensor), symmetrised with \irreps~$A_{2g}$ and $A_{2u}$, respectively}
\centering
\renewcommand{\arraystretch}{1.2}
\begin{tabular}{c|ccccccccc}
$c_{ijk}$ & 11 & 21 & 31 & 12 & 22 & 32 & 13 & 23 & 33 \\
\hline
1 & 0 & $c_{112}$ & $c_{131}$ & $c_{112}$ & 0 & 0 & $c_{113}$ & 0 & 0 \\
2 & $c_{112}$ & 0 & 0 & 0 & $-c_{112}$ & $c_{131}$ & 0 & $c_{113}$ & 0 \\
3 & $c_{311}$ & 0 & 0 & 0 & $c_{311}$ & 0 & 0 & 0 & $c_{333}$
\end{tabular}
\end{table}

\begin{table}[h]
\caption{\label{Tab: Gen_tensor_23} Cartesian intertwiner with Jahn symbols $V^2[V^3]$ (magneto-electric tensor) and $eV^2[V^3]$ (electrotoroidic tensor), symmetrised with \irreps~$A_{2u}$ and $A_{2g}$, respectively}
\centering
\resizebox{\textwidth}{!}{%
\renewcommand{\arraystretch}{1.2}
\begin{tabular}{c|*{18}{>{$}c<{$}}}
$c_{ijklm}$& 11 & 21 & 31 & 41 & 51 & 61 & 12 & 22 & 32 & 42 & 52 & 62 & 13 & 23 & 33 & 43 & 53 & 63 \\
\hline

11 &
0 & 0 & 0 & 0 &
c_{1113} & c_{1112} &
c_{1112} & c_{1122} & c_{1132} & c_{1123} &
0 & 0 &
c_{1113} & c_{1123} & c_{1133} & c_{1132} &
0 & 0
\\

21 &
\frac{3c_{1112}}2+\frac{3c_{1122}}2-c_{1211} &
-\frac{c_{1112}}2-\frac{c_{1122}}2+c_{1211} &
c_{1132} &
\frac{c_{1113}}2-\frac{c_{1123}}2 &
0 & 0 &
0 & 0 & 0 & 0 &
\frac{c_{1113}}2-\frac{c_{1123}}2 &
-\frac{c_{1112}}2-\frac{c_{1122}}2+c_{1211} &
0 & 0 & 0 & 0 &
c_{1132} &
\frac{c_{1113}}2-\frac{c_{1123}}2
\\

31 &
c_{3111} &
\frac{c_{3111}}3 &
c_{3131} &
c_{3141} &
0 & 0 &
0 & 0 & 0 & 0 &
c_{3141} &
\frac{c_{3111}}3 &
0 & 0 & 0 & 0 &
c_{3131} &
c_{3141}
\\

12 &
c_{1211} &
c_{1112}+c_{1122}-c_{1211} &
c_{1132} &
\frac{c_{1113}}2-\frac{c_{1123}}2 &
0 & 0 &
0 & 0 & 0 & 0 &
\frac{c_{1113}}2-\frac{c_{1123}}2 &
c_{1112}+c_{1122}-c_{1211} &
0 & 0 & 0 & 0 &
c_{1132} &
\frac{c_{1113}}2-\frac{c_{1123}}2
\\

22 &
0 & 0 & 0 & 0 &
c_{1123} &
\frac{c_{1112}}2-\frac{c_{1122}}2 &
\frac{c_{1112}}2-\frac{c_{1122}}2 &
-\frac{3c_{1112}}2-\frac{c_{1122}}2 &
-c_{1132} &
c_{1113} &
0 & 0 &
c_{1123} &
c_{1113} &
c_{1133} &
-c_{1132} &
0 & 0
\\

32 &
0 & 0 & 0 & 0 &
c_{3141} &
\frac{c_{3111}}3 &
\frac{c_{3111}}3 &
c_{3111} &
c_{3131} &
-c_{3141} &
0 & 0 &
c_{3141} &
-c_{3141} &
0 &
c_{3131} &
0 & 0
\\

13 &
c_{1311} &
\frac{c_{1311}}3 &
c_{1331} &
c_{1341} &
0 & 0 &
0 & 0 & 0 & 0 &
c_{1341} &
\frac{c_{1311}}3 &
0 & 0 & 0 & 0 &
c_{1331} &
c_{1341}
\\

23 &
0 & 0 & 0 & 0 &
c_{1341} &
\frac{c_{1311}}3 &
\frac{c_{1311}}3 &
c_{1311} &
c_{1331} &
-c_{1341} &
0 & 0 &
c_{1341} &
-c_{1341} &
0 &
c_{1331} &
0 & 0
\\

33 &
0 & 0 & 0 & 0 &
c_{3313} &
c_{3312} &
c_{3312} &
-c_{3312} &
0 &
c_{3313} &
0 & 0 &
c_{3313} &
c_{3313} &
c_{3333} &
0 & 0 & 0
\\

\end{tabular}
}
\end{table}

Likewise, converting the $V^2[V^3]$ intertwiner into the canonical form (canonical basis in $L^3$) one obtains for the three \irreps:

\noindent\textbf{\boldmath$A_{1g}$}
\begin{align}
\begin{pmatrix}
0 & A & 0 \\
-A & 0 & 0 \\
0 & 0 & 0
\end{pmatrix}.
\end{align}

\noindent\textbf{\boldmath$A_{2g}$}
\begin{align}
\begin{pmatrix}
B & 0 & 0 \\
0 & B & 0 \\
0 & 0 & C
\end{pmatrix},
\notag\\[3mm]
\begin{pmatrix}
D & 0 & 0 \\
0 & D & 0 \\
0 & 0 & E
\end{pmatrix},
\notag\\[3mm]
\begin{pmatrix}
F & 0 & 0 \\
0 & F & 0 \\
0 & 0 & G
\end{pmatrix}.
\end{align}

\noindent\textbf{\boldmath$E_g$}
\begin{align}
\begin{pmatrix}
0 & H & I \\
H & 0 & 0 \\
J & 0 & 0
\end{pmatrix}
\qquad
&
\begin{pmatrix}
H & 0 & 0 \\
0 & -H & I \\
0 & J & 0
\end{pmatrix},
\notag\\[3mm]
\begin{pmatrix}
0 & K & L \\
K & 0 & 0 \\
M & 0 & 0
\end{pmatrix}
\qquad
&
\begin{pmatrix}
K & 0 & 0 \\
0 & -K & L \\
0 & M & 0
\end{pmatrix},
\notag\\[3mm]
\begin{pmatrix}
0 & N & O \\
N & 0 & 0 \\
P & 0 & 0
\end{pmatrix}
\qquad
&
\begin{pmatrix}
N & 0 & 0 \\
0 & -N & O \\
0 & P & 0
\end{pmatrix}.
\end{align}

with 

\begin{align}
A &= -\frac{3}{4}c_{1112}-\frac{3}{4}c_{1122}+c_{1211}, &
B &= \frac{3}{4}c_{1112}-\frac{1}{4}c_{1122}, \notag\\
C &= c_{3312}, &
D &= \frac{3}{4}c_{1113}+\frac{3}{4}c_{1123}
     +\frac{1}{2}c_{1133}, \notag\\
E &= \frac{3}{2}c_{3313}+\frac{1}{2}c_{3333}, &
F &= \frac{3}{4}c_{1113}+\frac{3}{4}c_{1123}
     -\frac{1}{2}c_{1133}, \notag\\
G &= \frac{3}{2}c_{3313}-\frac{1}{2}c_{3333}, &
H &= \frac{3}{8}c_{1112}+\frac{3}{8}c_{1122}
     +\frac{3}{2}c_{1132}, \notag\\
I &= \frac{1}{2}c_{1311}+\frac{3}{2}c_{1331}, &
J &= \frac{1}{2}c_{3111}+\frac{3}{2}c_{3131}, \notag\\
K &= \frac{3}{8}c_{1112}+\frac{3}{8}c_{1122}
     -\frac{3}{2}c_{1132}, &
L &= \frac{1}{2}c_{1311}-\frac{3}{2}c_{1331}, \notag\\
M &= \frac{1}{2}c_{3111}-\frac{3}{2}c_{3131}, &
N &= \frac{3}{2}c_{1113}-\frac{3}{2}c_{1123}, \notag\\
O &= 3c_{1341}, &
P &= 3c_{3141}.
\end{align}

\section{ Magneto-electric tensor forms for Cr$_2$O$_3$ along high-symmetry directions }
\label{Sec: ME tensors}

All the tensors are expressed in trigonal coordinates, i.e., with the 2-fold axis along $x$ and the 3-fold axis along $z$.  Conversion to monoclinic coordinates requires an interchange of $x$ with $y$.

\begin{table}[h]
\caption{Magnetoelectric tensor for MPG $\bar{3}'2/m'$ ($\vec{L} \parallel c$) in the trigonal coordinate system, calculated using the programme MTENSOR from the Bilbao crystallographic server \cite{PerezMato2015, Gallego2019}}
\centering
\renewcommand{\arraystretch}{1.2}
\begin{tabular}{>{$}c<{$}|*{3}{>{$}c<{$}}}
\alpha_{ij} & 1 & 2 & 3 \\
\hline
1 & \alpha_{11} & 0 & 0\\
2 & 0 & \alpha_{11} &0 \\
3 &0 &  0 & \alpha_{33}
\end{tabular}
\end{table}

\begin{table}[h]
\caption{Magnetoelectric tensor for MPG $2'/m$ ($\vec{L} \parallel $ 2-fold axis) in the trigonal coordinate system, calculated using the programme MTENSOR from the Bilbao crystallographic server \cite{PerezMato2015, Gallego2019}}
\centering
\renewcommand{\arraystretch}{1.2}
\begin{tabular}{>{$}c<{$}|*{3}{>{$}c<{$}}}
\alpha_{ij} & 1 & 2 & 3 \\
\hline
1 & 0 & \alpha_{12} &  \alpha_{13}  \\
2 & \alpha_{21} & 0 &0\\
3 &  \alpha_{31} &0 & 0
\end{tabular}
\end{table}

\begin{table}[h]
\caption{Magnetoelectric tensor for MPG $2/m'$ ($\vec{L} \perp $ 2-fold axis) in the trigonal coordinate system, calculated using the programme MTENSOR from the Bilbao crystallographic server \cite{PerezMato2015, Gallego2019}}
\centering
\renewcommand{\arraystretch}{1.2}
\begin{tabular}{>{$}c<{$}|*{3}{>{$}c<{$}}}
\alpha_{ij} & 1 & 2 & 3 \\
\hline
1 & \alpha_{11} & 0 & 0\\
2 & 0 & \alpha_{22} & \alpha_{23}  \\
3 &0 &  \alpha_{32} & \alpha_{33}
\end{tabular}
\end{table}


\section{Computing general tensors using the fully canonical intertwiner method}
\label{App: Fully_Canonical_CG}

The canonical intertwiner framework I have described is extremely simple when mixed-variable canonical polynomials are employed: all one needs to do is to decompose the axial vector representation into \irreps~and write the corresponding canonical vectors.  The intertwiners simply consist of a list of copies of these vectors (with different parameters) for each instance of a given \irrep.  However, mixed-variable canonical polynomials make the physics rather opaque and are therefore generally inconvenient.   A much more convenient basis can be formed using products of single-variable canonical polynomials, such as $\Phi^{(\Gamma)}_{i, \alpha} (\vec{L}) \Xi^{(\Gamma')}_{j, \beta} (\vec{k})$ or even more complex polyadic forms involving three or more physical variables.  I shall call these generically `polyadic canonical bases'.   The advantage of these bases is that they make the harmonic angular dependence of each variable completely transparent.  Converting to such a basis is a simple matter of systematic application of Clebsch-Gordan coefficients.\newline

First of all, we can observe that different instances of the same \irrep~for a given variable never contribute to the same mixed-variable intertwiner.  Hence, the `instance labels' are simply tags that can be re-arranged at will.  Furthermore, Clebsch-Gordan coefficients only depend on the \irreps~and on the basis labels (Greek indices) and not on the specific instances. We can therefore write:

\begin{equation}
Z^{(\Gamma)}_{i, \alpha}= \sum_{\Gamma', \Gamma'', \beta, \gamma}C^{(\Gamma, \Gamma', \Gamma'')}_{\alpha, \beta,\gamma} \left[\Phi^{(\Gamma')}_{\beta} \Xi^{(\Gamma'')}_{\gamma} \right]_i
\end{equation}

without any need to specify which particular instance of $\Phi^{(\Gamma')}_{\beta}$ and $ \Xi^{(\Gamma'')}_{\gamma}$ we are considering, for example, in the list:

\begin{eqnarray}
\Phi^{(E_{g})}_{1,(2)} \Xi^{(E_g)}_{1,(1)} &=& \left[\Phi^{(E_{g})}_{2} \Xi^{(E_g)}_{1}\right]_1= L_y\!\left(L_x^2 + L_y^2 + L_z^2\right) \left(k_x^2-k_y^2\right)
\nonumber\\
\Phi^{(E_{g})}_{2,(2)} \Xi^{(E_g)}_{1,(1)} &=& \left[\Phi^{(E_{g})}_{2} \Xi^{(E_g)}_{1}\right]_2= L_y\!\left(L_x^2 + L_y^2 - L_z^2\right) \left(k_x^2-k_y^2\right)
\nonumber\\
\Phi^{(E_{g})}_{3,(2)} \Xi^{(E_g)}_{1,(1)} &=& \left[\Phi^{(E_{g})}_{2} \Xi^{(E_g)}_{1}\right]_3= L_z\!\left(L_x^2 - L_y^2\right)\left(k_x^2-k_y^2\right)
\nonumber\\
\Phi^{(E_{g})}_{1,(2)} \Xi^{(E_g)}_{2,(1)} &=& \left[\Phi^{(E_{g})}_{2} \Xi^{(E_g)}_{1}\right]_4= L_y\!\left(L_x^2 + L_y^2 + L_z^2\right) \left(k_yk_z\right)
\nonumber\\
\Phi^{(E_{g})}_{2,(2)} \Xi^{(E_g)}_{2,(1)} &=& \left[\Phi^{(E_{g})}_{2} \Xi^{(E_g)}_{1}\right]_5= L_y\!\left(L_x^2 + L_y^2 - L_z^2\right) \left(k_yk_z\right)
\nonumber\\
\Phi^{(E_{g})}_{3,(2)} \Xi^{(E_g)}_{2,(1)} &=& \left[\Phi^{(E_{g})}_{2} \Xi^{(E_g)}_{1}\right]_6= L_z\!\left(L_x^2 - L_y^2\right)\left(k_yk_z\right)
\end{eqnarray}

On the mixed-variable basis, we can write a generic tensor as:

\begin{equation}
\ten{T}=\sum_{\Gamma, i, \alpha} \vec{V}^{(\Gamma)}_{i,\alpha} Z^{(\Gamma)}_{i, \alpha}
\end{equation}

where we recall that the $\vec{V}^{(\Gamma)}_{i,\alpha} $ have the same canonical form for different $i$'s.  Hence, we can express the tensor as a function of a new set of vectors $\vec{W}^{(\Gamma', \Gamma'')}_{i,  \beta,\gamma}$ as:

\begin{equation}
\ten{T}=\sum_{ i} \vec{W}^{(\Gamma', \Gamma'')}_{i,  \beta,\gamma}\left[ \Phi^{(\Gamma')}_{\beta} \Xi^{(\Gamma'')}_{\gamma} \right]_i
\end{equation}

\begin{equation}
\vec{W}^{(\Gamma', \Gamma'')}_{i,  \beta,\gamma}=\sum_{\Gamma, i, \alpha} \vec{V}^{(\Gamma)}_{i,\alpha} C^{(\Gamma, \Gamma', \Gamma'')}_{\alpha, \beta,\gamma}
\end{equation}

where, once again, the $\vec{W}^{(\Gamma', \Gamma'')}_{i,  \beta,\gamma}$ have the same form regardless of the subscript $i$.\newline

From Equations \ref{Eq: l_cube_poly} and \ref{eq: Clebsch_Gordan}, one can immediately verify that:

\begin{eqnarray}
\vec{W}^{(A_{1g},A_{1g})}&=&\begin{pmatrix}
0\\
0 \\
a_1
\end{pmatrix}\nonumber\\
\vec{W}^{(A_{2g},A_{2g})}&=&\begin{pmatrix}
0\\
0 \\
a_2
\end{pmatrix}\nonumber\\
\vec{W}^{(A_{1g},E_g)}_{1}&=&\begin{pmatrix}
0\\
b_1 \\
0
\end{pmatrix},\qquad
\vec{W}^{(A_{1g},E_g)}_{2}=\begin{pmatrix}
-b_1\\
0 \\
0
\end{pmatrix}\nonumber\\
\vec{W}^{(A_{2g},E_g)}_{1}&=&\begin{pmatrix}
b_2\\
0 \\
0
\end{pmatrix}, \qquad
\vec{W}^{(A_{2g},E_g)}_{2}=\begin{pmatrix}
0\\
b_2 \\
0
\end{pmatrix}\nonumber\\
\vec{W}^{(E_g, A_{1g})}_{1}&=&\begin{pmatrix}
0\\
b_3 \\
0
\end{pmatrix},\qquad
\vec{W}^{(E_g, A_{1g},)}_{2}=\begin{pmatrix}
-b_3\\
0 \\
0
\end{pmatrix}\nonumber\\
\vec{W}^{(E_g, A_{2g})}_{1}&=&\begin{pmatrix}
b_4\\
0 \\
0
\end{pmatrix}, \qquad
\vec{W}^{(E_g, A_{2g},)}_{2}=\begin{pmatrix}
0\\
b_4 \\
0
\end{pmatrix}\nonumber\\
\vec{W}^{(E_g,E_g)}_{1,(1)}&=&\begin{pmatrix}
0\\
b_5 \\
a_3
\end{pmatrix}, \qquad
\vec{W}^{(E_g,E_g)}_{1,(2)}=\begin{pmatrix}
b_5\\
0 \\
0
\end{pmatrix}\nonumber\\
\vec{W}^{(E_g,E_g)}_{2,(1)}&=&\begin{pmatrix}
b_5\\
0 \\
0
\end{pmatrix}, \qquad
\vec{W}^{(E_g,E_g)}_{2,(2)}=\begin{pmatrix}
0\\
-b_5 \\
a_3
\end{pmatrix}
\end{eqnarray}

For two physical variables (e.g., the order parameter $\vec{L}$ and the momentum $\vec{k}$), these vectors are all that is required to construct canonical intertwiners at any order on the dyadic basis --- for example $\Phi^{(\Gamma)}_{i, \alpha} (\vec{L}) \Psi^{(\Gamma')}_{j, \beta} (\vec{k})$.  For multiple variables, one can construct the appropriate vector forms (such as $\vec{W}^{(\Gamma', \Gamma'', \Gamma''')}_{\beta,\gamma, \delta}$ for three variables) iteratively using the same Clebsch-Gordan coefficients.\newline

These considerations naturally suggest a computational approach to constructing general intertwiners expressed in terms of polyadic canonical bases:

\begin{description}
\item[Routine 1] Constructs the canonical vector forms $\vec{V}^{(\Gamma)}$ given the parent group (one of the 32 crystal classes) and one of its 1D \irreps.  
\item[Routine 2] Constructs a library of Clebsch-Gordan coefficients $C^{(\Gamma, \Gamma', \Gamma'')}_{\alpha, \beta,\gamma}$ for each of the 32 crystal classes. 
\item[Routine 3] Constructs a library of symmetry-adapted single-variable polynomials at any order for each of the 32 crystal classes.
\item[Routine 4] Constructs iteratively the $\vec{W}^{(\Gamma', \Gamma'', \dots)}_{\beta, \gamma, \dots}$ vector forms given the Clebsch-Gordan coefficients and a generic set of canonical vector forms $\vec{V}^{(\Gamma)}_\alpha$.  
\item[Routine 5] Reassembles the tensor, given its Jahn symbol.  Each polyadic term, e.g., $\Phi^{(\Gamma')}_{i,\alpha}(\vec{L}) \Psi^{(\Gamma'')}_{j,\beta}(\vec{k}) \Upsilon^{(\Gamma''')}_{k,\gamma}(\vec{E})\dots$ will get a copy of the $\vec{W}^{(\Gamma', \Gamma'', \dots)}_{\alpha, \beta, \dots}$ with unique coefficients.  There is no need to track the $i,j,k, \dots$ as long as the coefficients are unique.
\end{description}

Routine 1 is the only one that employs the bipartition \irrep, and therefore depends on magnetic ordering.  Routines 2, 3 and 4 can construct small libraries once and for all.  Routine 5 is completely generic and does not depend on the crystal class, bipartition \irrep~or Jahn symbol, so long as the appropriate parameters are provided as inputs.

\newpage
\begin{acknowledgements}
The author would like to acknowledge discussions with Paul Zeng (Oxford, Hamburg) and Hariom K. Jani (Oxford).
\end{acknowledgements}

\bibliography{Altermag_Irrep_2026.bib}

\end{document}